\documentclass[aps,prl,twocolumn,superscriptaddress,nofootinbib]{revtex4-1} 

\usepackage{chemformula} 
\usepackage{longtable} 
\usepackage{multirow} 
\usepackage{amssymb} 

\usepackage{braket} 

\usepackage{amsmath} 

\usepackage{float}

\usepackage[figuresright]{rotating} 

\usepackage{cfr-lm}

\usepackage{xr}

\begin{document}

\title{Structure-Imposed Electronic Topology in Cove-Edged Graphene Nanoribbons}

\author{Florian M. Arnold}
    \altaffiliation{Equal contribution}
    \affiliation{Faculty of Chemistry and Food Chemistry, TU Dresden, Bergstrasse 66c, 01069 Dresden, Germany}
\author{Tsai-Jung Liu}
    \altaffiliation{Equal contribution}
    \affiliation{Faculty of Chemistry and Food Chemistry, TU Dresden, Bergstrasse 66c, 01069 Dresden, Germany}
\author{Agnieszka Kuc}
    \affiliation{Helmholtz-Zentrum Dresden-Rossendorf, Institute of Resource Ecology, Permoserstrasse 15, 04318 Leipzig, Germany}
\author{Thomas Heine}
    \email{thomas.heine@tu-dresden.de}
    \affiliation{Faculty of Chemistry and Food Chemistry, TU Dresden, Bergstrasse 66c, 01069 Dresden, Germany}
    \affiliation{Helmholtz-Zentrum Dresden-Rossendorf, Institute of Resource Ecology, Permoserstrasse 15, 04318 Leipzig, Germany}
    \affiliation{Department of Chemistry, Yonsei University, Seodaemun-gu, Seoul 120-749, Republic of Korea}

\date{\today}

\begin{abstract}
In cove-edged zigzag graphene nanoribbons (ZGNR-Cs), one terminal \ch{CH} group per length unit is removed on each zigzag edge, forming a regular pattern of coves that controls their electronic structure.
Based on three structural parameters that unambiguously characterize the atomistic structure of ZGNR-Cs, we present a scheme that classifies their electronic state, i.e., if they are metallic, topological insulators or trivial semiconductors, for all possible widths $N$, unit lengths $a$ and cove position offsets at both edges $b$, thus showing the direct structure-electronic structure relation.
We further present an empirical formula to estimate the band gap of the semiconducting ribbons from $N,$ $a$, and $b$. 
Finally, we identify all geometrically possible ribbon terminations and provide rules to construct ZGNR-Ca with a well-defined electronic structure.
\end{abstract}

\maketitle



\begin{figure}[ht!]
    \centering
    \includegraphics[width=0.9\columnwidth]{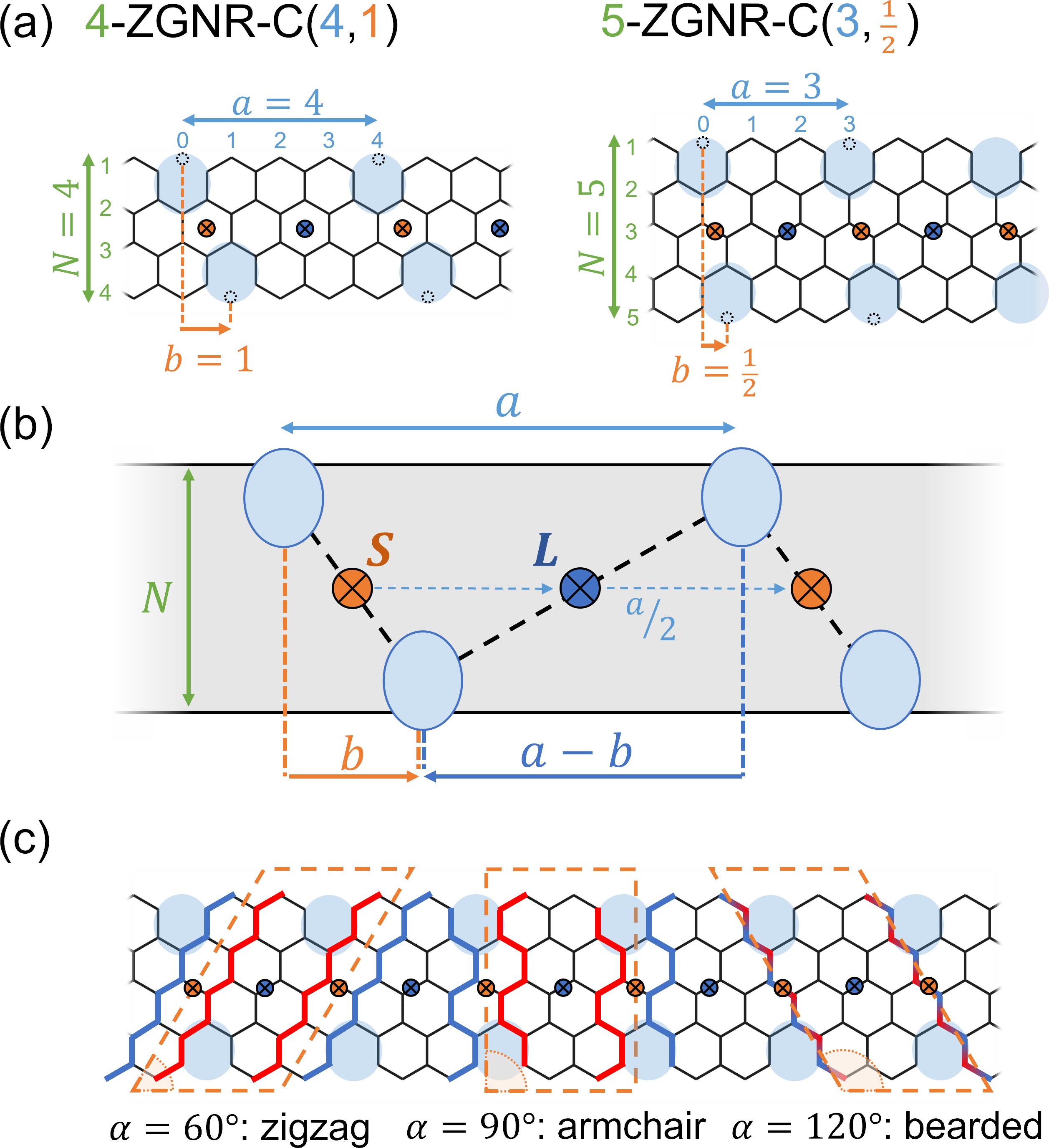}
    \caption{(a) Visualization of parameters $N$, $a$ and $b$ in $N$-ZGNR-C($a$,$b$). (b) Schematic representation of parameters and positions of reference points \textit{\textbf{S}} and \textit{\textbf{L}} relative to cove positions. (c) Termination types (red and blue) at boundary of unit cell (orange) for different values of $\alpha$ in 5-ZGNR-C(3,$\frac{1}{2}$).}
    \label{fig_nomenclature}
\end{figure}

With the emergence of precision chemistry, the atomically precise synthesis of graphene nanoribbons (GNRs) with well-defined widths, edge structures, and terminations has become 
possible~\cite{ruffieux2016surface,houtsma2021atomically,yoon2020liquid,narita2019solution}.
Besides GNRs with zigzag or armchair edge termination, more complex edges have also been realized~\cite{groning2018engineering,rizzo2018topological,yoon2020liquid,narita2019solution,wang2021coves,shinde2021graphene,yao2021synthesis,li2021topological,hu2018bandgap,cai2014graphene,houtsma2021atomically,liu2015toward}.
The edges have determinant impact on the electronic structure of the GNRs.
In their seminal work, Lee et al.~\cite{lee2018coves} have demonstrated topologically nontrivial cove-edged zigzag GNRs (ZGNR-Cs) and the influence of cove placement and ribbon width.
The introduction of coves -- missing CH groups at both edges of an inherently metallic zigzag GNR, forming a superlattice -- opens a band gap.
ZGNR-Cs have been realized experimentally~\cite{liu2015toward,wang2021coves}; but, up to now their topological states have been predicted only by theory~\cite{liu2020interface}.
Although many of the ZGNR-Cs have been studied to date, there is not yet any rule established that connects their structural topology with their electronic state.
Such a connection is well known and extremely useful for carbon nanotubes, where the chirality indexes $(m,n)$ define the tube's electronic state, being either metallic if $(m-n)/3$ is an integer, or semiconducting otherwise~\cite{hamada1992new,saito1992electronic_C60,saito1992electronic}.
For armchair GNRs, the dependence of the topological invariant on the width and unit cell is known~\cite{cao2017topological}.
Here, we report the direct relation between three structural parameters that identify ZGNR-Cs (see Fig.~\ref{fig_nomenclature}) and the electronic state, and we discuss the impact of termination effects.

Unambiguous labeling of cove-edged ZGNRs requires three parameters: the width $N$, counted as the number of zigzag rows of carbon atoms ($N\in \mathbb{N}, N\geq 3$); the distance between coves on the same edge $a$ in units of hexagonal rings ($a\in\mathbb{N}, a\geq 2$), which also gives the unit cell length; and the shortest offset between adjacent coves on opposite edges $b$ ($b\in\left[ 0, \sfrac{a}{2} \right]$). 
The $N$-ZGNR-C($a,b$), illustrated in Figs.~\ref{fig_nomenclature}(a) and \ref{fig_nomenclature}(b), labels all structurally possible regular ZGNR-Cs and is equivalent to the $N$-CZGNR-($m$,$n$) introduced by Wang et al. with $a=m+1$~\cite{wang2021coves}.
For even $N$, $b$ is an integer; whereas for odd $N$, $b$ is a half-integer.
These parameters directly relate to the Frieze groups~\cite{IUC_Frieze} as $p2mm$ for $b=0$, $p2mg$ for $b=\sfrac{a}{2}$, and $p211$ otherwise.
This is shown for exemplary structures in Fig.~S1 in the Supplemental Material (SM)~\cite{SM}.

For a given ZGNR-C with $b\neq\sfrac{a}{2}$, there are two distinct inversion centers in the unit cell, henceforth called reference points \textit{\textbf{S}} (denoting the inversion center at the shorter offset between the coves) and \textit{\textbf{L}} (located at the longer offset), as shown in  Fig.~\ref{fig_nomenclature}(b).
\textit{\textbf{S}} and \textit{\textbf{L}} are shifted by half a lattice vector along the ribbon with respect to one another.
For $b=\sfrac{a}{2}$, \textit{\textbf{S}} and \textit{\textbf{L}} are equivalent.
In general, the position of the unit cell of any structure is arbitrary. 
Here, we concentrate on GNRs as obtained in bottom-up synthesis from well-defined monomers. 
Typically, either one centrosymmetric or two identical building blocks are used~\cite{houtsma2021atomically}.
As a consequence, the unit cell becomes centrosymmetric and has inversion centers at the boundaries.
Illustrative examples can be found in published bottom-up syntheses of ZGNR-Cs~\cite{liu2015toward,wang2021coves}.
The molecular building blocks determine not only the ribbon geometry but also its terminal ends.
Therefore, cell angles of $\alpha=60$, $90$, and $120^\circ$ are chosen because they coincide with the carbon-carbon bond orientation.
The unit cell boundary then cuts through the ribbon in different ways, creating either an armchair, zigzag or bearded termination type (see Fig.~\ref{fig_nomenclature}(c)).

The tight-binding (TB) method is applied to calculate electronic properties and topological invariants with the PythTB package~\cite{PythTB} (see SM Sec.~S1 for details).
By considering one $\pi$ electron per carbon atom, the TB model Hamiltonian with only first-neighbor interactions $t_1=-1$ and on-site energies $\varepsilon_{i}$ is
\begin{equation}
H=\sum_{i}{\varepsilon_{i}c^{\dagger}_{i}c_{i}}+\sum_{\left\langle{i,j}\right\rangle}{t_{1}c^{\dagger}_{i}c_{j}}.
\end{equation}
To characterize the topological properties, the Zak phase~\cite{10.1103/PhysRevLett.62.2747} was calculated as an integral of the Berry connection $i\braket{u_{n\boldsymbol{k}}|\nabla_{\boldsymbol{k}} u_{n\boldsymbol{k}}}$ over the Brillouin zone (BZ) summed over all occupied states,
\begin{equation}
\gamma=i\sum_{n=1}^{\text{occ.}}\int_C\braket{u_{n\boldsymbol{k}}|\nabla_{\boldsymbol{k}} u_{n\boldsymbol{k}}}\cdot d\boldsymbol{k},
\end{equation}
where $C$ is an open path over the BZ.
This can be divided into two parts: 1) integrating within the first BZ, and 2) crossing the boundary from the last point of the cell to the first point of the next periodic cell.
$\gamma$ is then calculated numerically as in \cite{resta2000manifestations} using the gauge-periodic-boundary condition,
\begin{equation}
u_{n\boldsymbol{k}_\mathrm{final}}=e^{-i\boldsymbol{G}\boldsymbol{r}}u_{n\boldsymbol{k}_\mathrm{initial}},    
\end{equation}
with reciprocal-lattice vector $\boldsymbol{G}$.
For systems with inversion or mirror symmetries in the unit cell, the Zak phase is quantized to zero or $\pi$ when one of the inversion centers coincides with the real-space coordinate origin~\cite{rhim2017bulk} (as it is the case for all systems studied here), and the$\mathbb{Z}_2$ topological invariant is obtained from 
\begin{equation}
    \mathbb{Z}_2 = \left( \frac{\gamma}{\pi} \right) \mathrm{mod}\ 2.
\end{equation}

Following Fu and Kane~\cite{10.1103/PhysRevB.76.045302}, $\mathbb{Z}_2$ values for systems with inversion symmetry can also be obtained from the parity of occupied states at time-reversal invariant momentum (TRIM) points as
\begin{equation}
    (-1)^{\mathbb{Z}_2} = \prod_m \prod_n^{\text{occ.}} \xi(\psi_n),
\end{equation}
where $m$ are the TRIM points, and $\xi(\psi_n)$ is the parity of the $n$th occupied band.

The topological signatures obtained by the Zak phase and parity calculations agree for all our systems.
For systems without inversion symmetry alternative methods are available~\cite{lin2018topological,jiang2020topology}. However, motivated by the available experiments we restrict our analyses to symmetric ZGNR-Cs.

\begin{figure*}[ht!] 
    \centering
    \includegraphics[width=\textwidth]{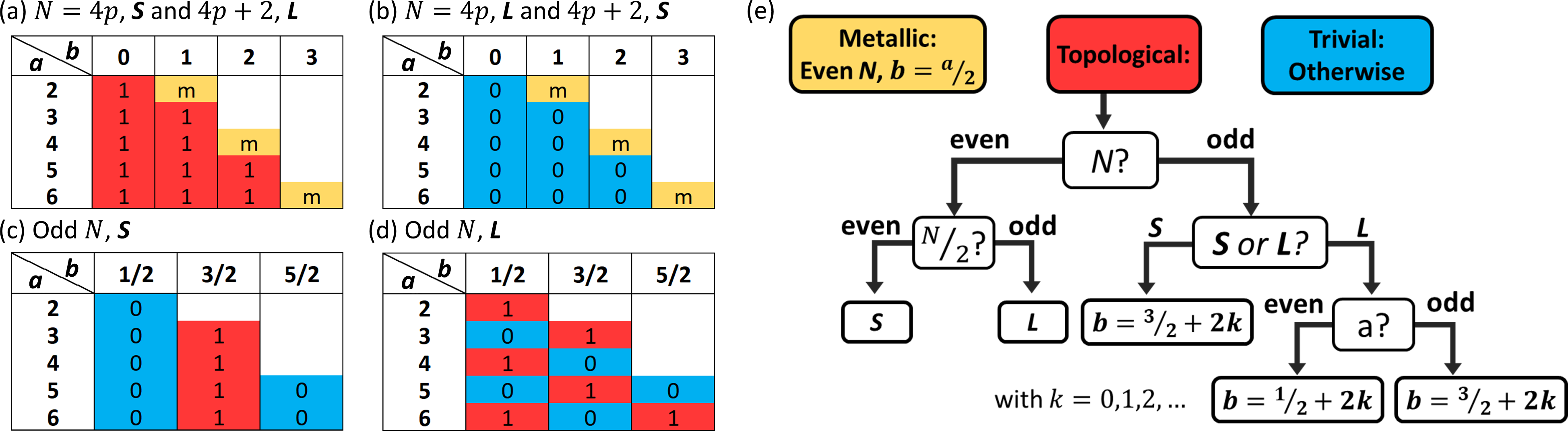}
     \caption{$\mathbb{Z}_2$ invariant for ZGNR-C in (a) $N=4p$ ($p = 1,2,\ldots$) with \textit{\textbf{S}} or $N=4p+2$ with \textit{\textbf{L}} at the boundary, (b) $N=4p$ with \textit{\textbf{L}} or $N=4p+2$ with \textit{\textbf{S}} at the boundary, (c) $N=4p+1$ and $N=4p+3$ with \textit{\textbf{S}} and (d) $N=4p+1$ and $N=4p+3$ with \textit{\textbf{L}} at the boundary. Structural parameters $a$ and $b$ are varied in rows and columns, respectively. Topological insulators ($\mathbb{Z}_2=1$) are marked in red, trivial semiconductors ($\mathbb{Z}_2=0$) in blue, and metallic ribbons (marked ``m") in yellow. (e) Generalized scheme to derive $\mathbb{Z}_2$ for a given set of structural parameters $N$, $a$, and $b$.}
    \label{fig_Z2_rules}
\end{figure*}

By sampling the configuration space of cove-edged ZGNRs, we find that the structural parameters $N$, $a$, and $b$ determine the ribbon's electronic state to be either metallic or semiconducting, as well as their $\mathbb{Z}_2$ invariant.
Using a classification based on the reference points \textit{\textbf{S}} and \textit{\textbf{L}} at the unit cell boundary, we observe that $\mathbb{Z}_2$ is uniquely given by the structural parameters, with a 4$p$ periodicity in $N$ ($p = 1,2,\ldots$) as shown for small $a$ in Figs.~\ref{fig_Z2_rules}(a)-\ref{fig_Z2_rules}(d) and extended to large $a$ in Fig.~S2.
A classification scheme of the electronic character of cove-edged ZGNR structures is given in Fig.~\ref{fig_Z2_rules}(e).

For even $N$, the topological properties only depend on the reference point at the unit cell boundary, independent of $a$ and $b$, with the exception of the metallic ZGNR-C with $b=\sfrac{a}{2}$ [Figs.~\ref{fig_Z2_rules}(a) and \ref{fig_Z2_rules}(b)].
The $N=4p \, {(p=1,2,\ldots)}$ ZGNR-C with \textit{\textbf{S}} and the ${N=4p+2}$ ZGNR-C with \textit{\textbf{L}} at the boundary are topologically nontrivial, whereas the others are trivial semiconductors.
This can be reflected in the expression ${\mathbb{Z}_2=\left[ \left(\sfrac{N}{2}\right) +1 \right] \,\mathrm{mod}\ 2}$ for \textit{\textbf{S}} and $\mathbb{Z}_2=\left( \sfrac{N}{2} \right) \,\mathrm{mod}\ 2$ for \textit{\textbf{L}}.
For odd $N$, the same values of $\mathbb{Z}_2$ are obtained for $N=4p+1$ and $N=4p+3$ with the same reference point.
The topological properties become independent of $a$ and only change with the value of $b$ as $\mathbb{Z}_2=(b-\frac{1}{2})\,\mathrm{mod}\ 2$ for reference point \textit{\textbf{S}} [Fig.~\ref{fig_Z2_rules}(c)].
For \textit{\textbf{L}}, the $\mathbb{Z}_2$ invariant interchanges with both $a$ and $b$ as $\mathbb{Z}_2=(a+b+\frac{1}{2})\,\mathrm{mod}\ 2$ in a checkerboard pattern [Fig.~\ref{fig_Z2_rules}(d)].
These rules can be generalized by a single function for $\mathbb{Z}^{\boldsymbol{S}}_2$ as the value of $\mathbb{Z}_2$ for reference point \textit{\textbf{S}}, which is given by
\begin{equation}
\begin{split}
    \label{eq_rules_Z2_S}
    \mathbb{Z}^{\boldsymbol{S}}_2= & \left(N+1\right)\mathrm{mod}\ 2 \cdot \left(\frac{N}{2}+1\right)\mathrm{mod}\ 2 \\
    & + \left(N\right)\mathrm{mod}\ 2 \cdot \left(b-\frac{1}{2}\right)\mathrm{mod}\ 2,
\end{split}
\end{equation}
and a function relating $\mathbb{Z}^{\boldsymbol{S}}_2$ to $\mathbb{Z}^{\boldsymbol{L}}_2$ ($\mathbb{Z}_2$ for reference point \textit{\textbf{L}}), which is given by
\begin{equation}
    \label{eq_rules_Z2_L}
    \mathbb{Z}^{\boldsymbol{L}}_2=|\left(Na+1\right)\mathrm{mod}\ 2- \mathbb{Z}^{\boldsymbol{S}}_2|.
\end{equation}

\begin{figure}[ht!]
    \centering
    \includegraphics[width=\columnwidth]{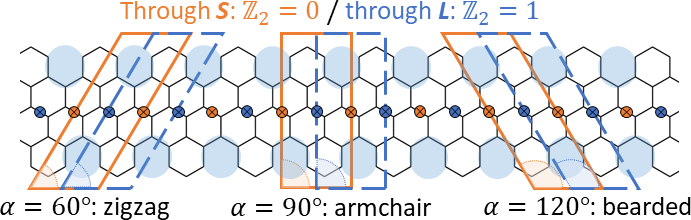}
    \caption{Possible primitive unit cells for 5-ZGNR-C(2,$\frac{1}{2}$) ribbons with reference points  \textit{\textbf{S}} (orange) and \textit{\textbf{L}} (blue) at the boundary. Unit cells with $\alpha=60^\circ$, $\alpha=90^\circ$, and $\alpha=120^\circ$ are shown; and the termination type is indicated.}
    \label{fig_termination}
\end{figure}

The value of $\mathbb{Z}_2$ does not depend on the termination type of the unit cell; however, the positions of \textit{\textbf{S}} and \textit{\textbf{L}} (Fig.~\ref{fig_nomenclature}) can affect $\mathbb{Z}_2$.
As shown in Fig.~\ref{fig_termination}, for the unit cells of a given $N$-ZGNR-C($a$,$b$) having the same reference point at the boundary, the same value of $\mathbb{Z}_2$ is obtained.
It is independent of the realized termination type being an armchair, a zigzag or bearded.

The change of $\mathbb{Z}_2$ upon exchange of \textit{\textbf{S}} and \textit{\textbf{L}} at the unit cell boundary is apparent from the parity of the electronic states.
In the TB formalism, the eigenstate $\psi_n$ is expressed as 
\begin{equation}
    \psi_n=\sum_i c_{n,i}e^{i\boldsymbol{k}\boldsymbol{r}_i}| \phi_i\rangle = \sum_i \tilde{c}_{n,i}| \phi_i\rangle,
\end{equation}
where $c_{n,i}$ are the components $i$ of eigenvector $\boldsymbol{c}_n$ to the basis functions $\phi_i$, and $\boldsymbol{r}$ is the position of site $i$.
When treated as a periodic infinite system, the same ZGNR-C will always give the same set of eigenvectors $\boldsymbol{c}_n$.
The parity is only affected by the structure of the ribbon itself and by the choice of its terminal position (\textit{\textbf{S}} or \textit{\textbf{L}} at the boundary).
When switching the boundary between reference points \textit{\textbf{S}} and \textit{\textbf{L}}, the unit cell is shifted by half a lattice vector.
At $k=0$, this transformation has no effect because $e^{i\boldsymbol{k}\boldsymbol{r}_i}=1$.
On the other hand, at $k=\pi$, half of the elements in $\tilde{c}_{n,i}$ are multiplied by $-1$, and parity switches the sign for all states at $k=\pi$.
As the number of occupied states in a ZGNR-C is $Na-1$, $\mathbb{Z}_2$ interchanges when the boundary of the unit cell is shifted between \textit{\textbf{S}} and \textit{\textbf{L}} for all structures with an odd number of occupied states as their parity product changes sign at $k=\pi$.
Hence, $\mathbb{Z}_2$ stays the same when switching between \textit{\textbf{S}} and \textit{\textbf{L}} only for ZGNR-Cs with both odd $N$ and odd $a$ at the same time due to an even number of occupied states.
Similar parity-based arguments rationalize why the cutting angle $\alpha$, and thus the structure of the terminals (zigzag, armchair, or bearded), does not affect the value of $\mathbb{Z}_2$.

\begin{figure}[ht!]
    \centering
    \includegraphics[width=\columnwidth]{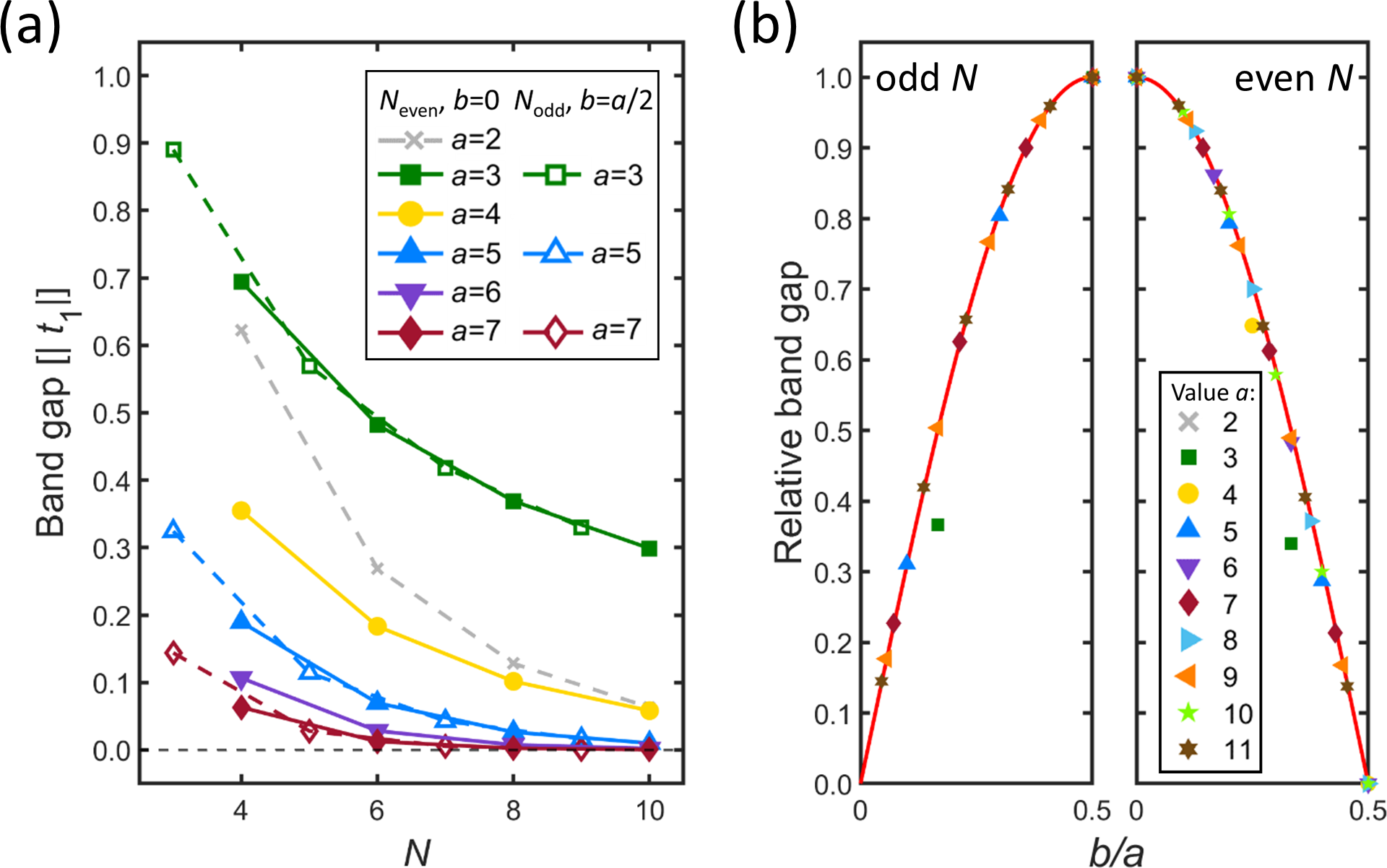}
    \caption{(a) $\Delta_\mathrm{max}$ in units of $|t_1|$. (b) $\Delta_\textui{g}$ relative to the maximum value for a given $a$ as a function of $b/a$ for (left) odd and (right) even $N$. Data points correspond to calculated structures, and continuous red lines correspond to the fit of Eq.~(\ref{eq_band_gap}).}
    \label{fig_band_gap}
\end{figure}

The parameters $N$, $a$, and $b$ also define the character and size of the band gap $\Delta_\textui{g}$.
All semiconducting ZGNR-Cs have a direct band gap at $k=0$ for odd $a$ and at $k=\pi$ for even $a$, independent of $N$ and $b$ (see Fig.~S3 for examples).
Quantum confinement gives the largest $\Delta_\textui{g}$ for small $N$, and $\Delta_\textui{g}$ decreases exponentially with $N$ [see Fig.~\ref{fig_band_gap}(a)].
This was also reported for other GNR types~\cite{saroka2014edge}.
We also observe exponential decay with increasing $a$ because the cove is the structural element opening the band gap in zigzag nanoribbons.
The only exception is found for $a=2$, where the proximity of the coves dominates the electronic structure.
Because of this, the corresponding data points in Fig.~\ref{fig_band_gap}(a) are colored in gray.

The dependence of $\Delta_\textui{g}$ on offset $b$, expressed relative to $a$, is shown in Fig.~\ref{fig_band_gap}(b).
The largest $\Delta_\textui{g}$ values for a given $a$ ($\Delta_\mathrm{max}$) are found at $b=0$ for even $N$ and at $b=\sfrac{a}{2}$ for odd $N$.
As a consequence, ZGNR-Cs with even $N$ are metallic for $b=\sfrac{a}{2}$.
However, geometric out-of-plane distortion (e.g. by H-H repulsion across the cove) reduces the $p2mg$ symmetry and opens a band gap~\cite{wang2021coves,son2006energy}.
These results show that ZGNR-Cs with large band gaps are obtained for thin ribbons (small $N$), for high cove density (small $a$), and for $b$ being close to zero for even or to $\sfrac{a}{2}$ for odd $N$.

The relative band gap as a function of $\sfrac{b}{a}$ converges toward a cosine function [see Fig.~\ref{fig_band_gap}(b)].
The deviations evident for $a=3$ quickly decrease with increasing $a$.
Thus, a good estimate for $\Delta_\textui{g}$ for all ZGNR-Cs is
\begin{equation}
    \Delta_\textui{g}= \Delta_\mathrm{max} \cdot \cos \left[\left(\frac{b}{a}+\frac{N\mathrm{mod\ 2}}{2}\right)\pi\right],
    \label{eq_band_gap}
\end{equation}
where $\Delta_\mathrm{max}$ can be fitted as function of $N$ and $a$ ($a\geq3$):
\begin{equation}
    \Delta_\mathrm{max}=Ae^{-B\cdot N}+Ce^{-(D+E\cdot a)\cdot N}.
    \label{eq_Delta_max}
\end{equation}
Close agreement with explicit TB calculations (${\mathrm{R}^2>0.99}$) is obtained for $A=3.04$, $B=1.42$, $C=1.30$, $D=-0.35$, and $E=0.17$ (see Fig.~S4(a)), parameterized for data with $3\leq N\leq 10$ and $3\leq a\leq 11$.
For large values of $N$ with $a=3$, the convergence of $\Delta_\textui{g}$ toward zero is slower than predicted by Eq.~(\ref{eq_Delta_max}) because of short-range effects [see Fig.~S4(b)].

\begin{figure}[ht!]
    \centering
    \includegraphics[width=0.45\textwidth]{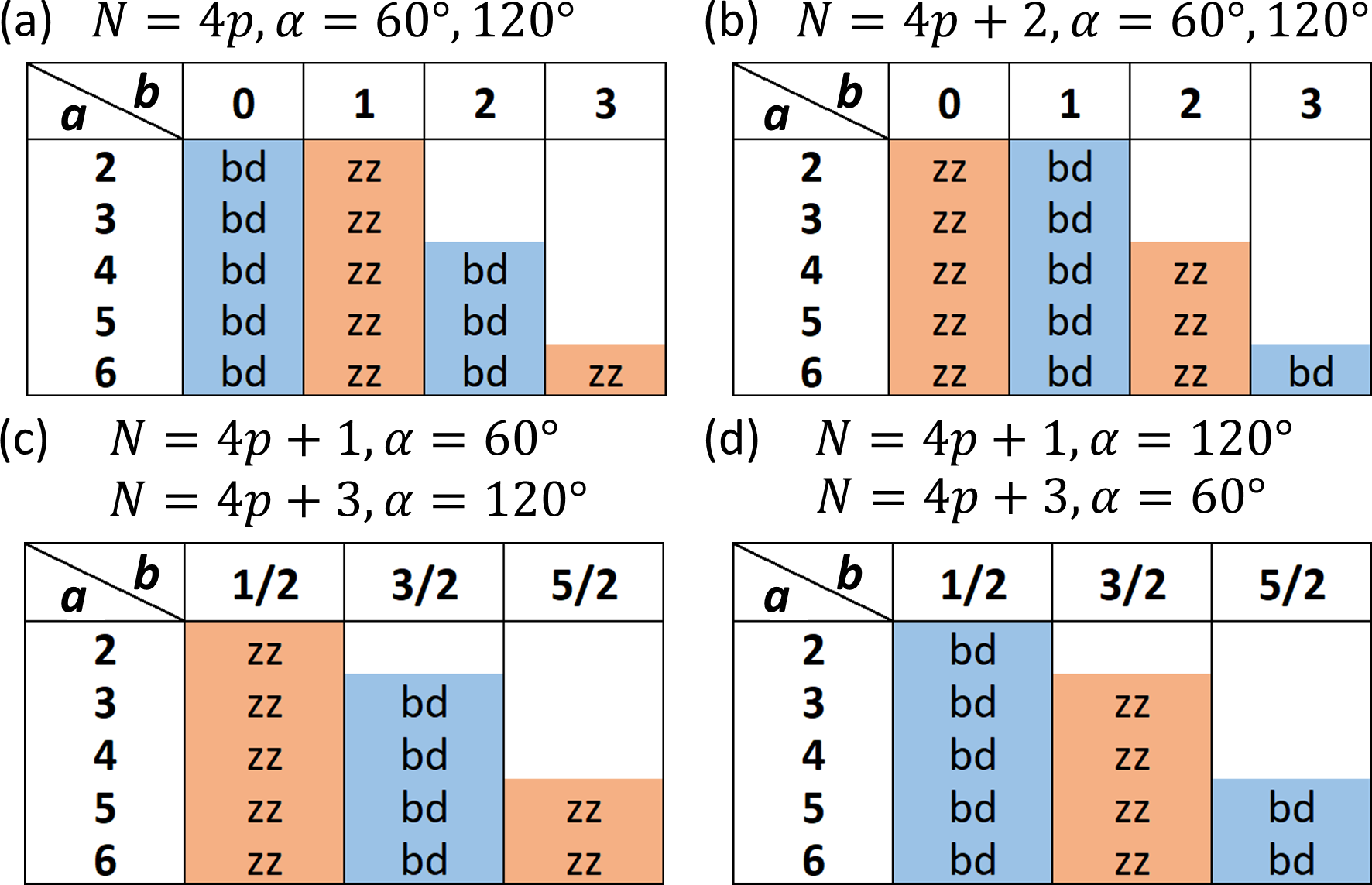}
    \caption{Realized termination types for ZGNR-C with \textit{\textbf{S}} at the unit cell boundary: (a) ${N=4p}$ and (b) ${N=4p+2}$ show the same termination types for both ${\alpha=60^\circ}$ and ${\alpha=120^\circ}$. (c) Termination types in ${N=4p+1}, {\alpha=60^\circ}$ and ${N=4p+3}, {\alpha=120^\circ}$; and (d) termination types in ${N=4p+1}, {\alpha=120^\circ}$ and ${N=4p+3}, {\alpha=60^\circ}$. Zigzag terminations are indicated by "zz" in orange, and bearded  terminations are indicated by "bd" in blue.}
    \label{fig_termination_rules}
\end{figure}

A junction state occurs when a topologically trivial and a topologically nontrivial GNR fuse. This requires commensurable terminal ends, which can be expressed by commensurable unit cell boundaries.
The termination types in ZGNR-Cs can be described by simple rules as shown in Fig.~\ref{fig_termination_rules} for reference point \textit{\textbf{S}}, and in Fig.~S5 for both \textit{\textbf{S}} and \textit{\textbf{L}} with an extended dataset.
A schematic representation similar to Fig.~\ref{fig_Z2_rules}(e) is shown in Fig.~S6.

All rectangular unit cells have armchair terminations.
However, for even $N$, the symmetry along the ribbon axis is broken in finite ribbons due to atoms at the boundary of the cell.

In ZGNR-Cs with even $N$, either zigzag or bearded termination is realized for both $\alpha=60^\circ$ and $\alpha=120^\circ$.
The influence of $b$ on the termination type is shown in Figs.~\ref{fig_termination_rules}(a) and \ref{fig_termination_rules}(b) for \textit{\textbf{S}} and in Figs.~S5(e) and S5(f) for \textit{\textbf{L}}.
For \textit{\textbf{S}}, a bearded termination is found for ${b=2k}$ ($k=0,1,2,\ldots$) and a zigzag termination is found for ${b=1+2k}$ at $N=4p$.
This interchanges for ${N=4p+2}$.
For \textit{\textbf{L}}, the zigzag and bearded terminations alternate with both $a$ and $b$ for both $N=4p$ and $N=4p+2$.

In ZGNR-Cs with odd $N$, both zigzag and bearded terminations are realized by choosing either $\alpha=60^\circ$ or $\alpha=120^\circ$, as shown in Figs.~\ref{fig_termination_rules}(c) and \ref{fig_termination_rules}(d) and Figs.~S5(c)-S5(f).
With \textit{\textbf{S}} at the unit cell boundary, the termination types in ZGNR-Cs with odd $N$ are independent of $a$ and alternate with $b$.
Zigzag terminations are realized by $\alpha=60^\circ$ at even $b-\frac{1}{2}$ with $N=4p+1$ and at odd $b-\frac{1}{2}$ with $N=4p+3$. 
In cells with $\alpha=120^\circ$, zigzag and bearded terminations are interchanged.
With \textit{\textbf{L}} at the unit cell boundary, the realized termination type alternates with both $a$ and $b$.
Detailed information on this is given in Figs.~S5(e) and S5(f).

\begin{figure}[ht!]
    \centering
    \includegraphics[width=0.45\textwidth]{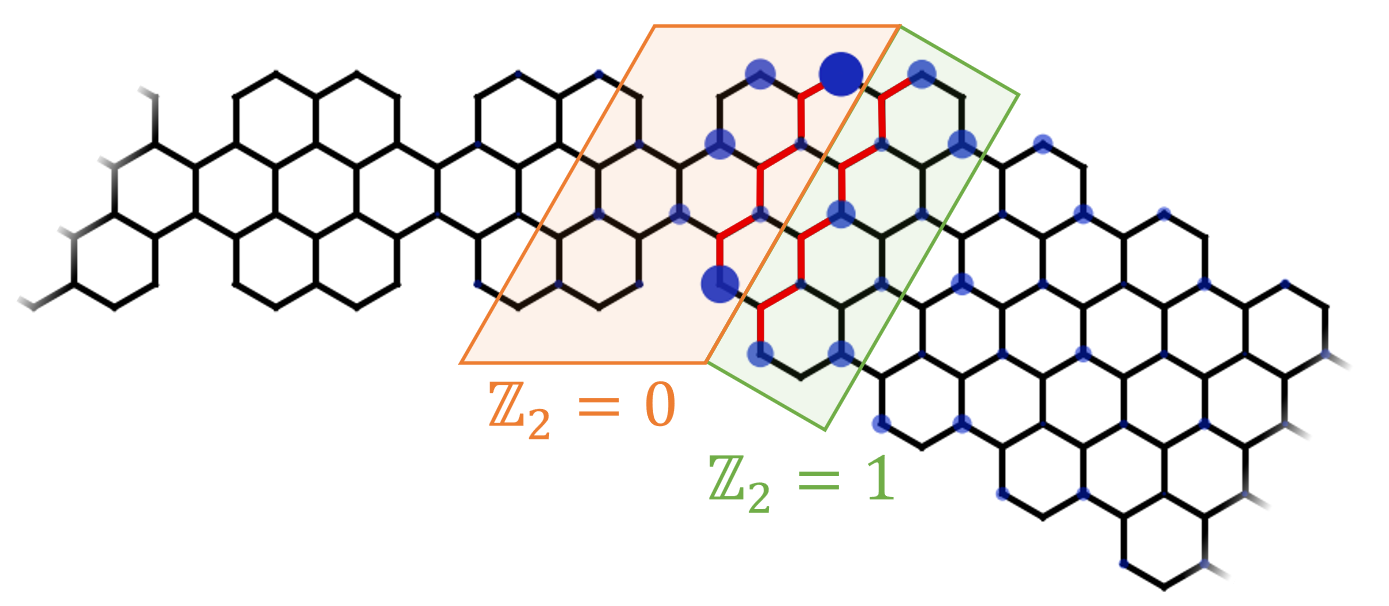}
    \caption{Topological junction state between 4-ZGNR-C(3,0) with reference point \textit{\textbf{L}} at the boundary of the unit cell (${\mathbb{Z}_2=0}$) and pristine 9-AGNR of the zigzag cell type (${\mathbb{Z}_2=1}$). The ribbons are connected at a zigzag termination (red line), and the primitive unit cells adjacent to the heterojunction are indicated.}
    \label{fig_edge_state}
\end{figure}

Together with the rules on the $\mathbb{Z}_2$ invariant, the band gap, and finite ZGNR-C structures, we have now developed the rule set for the construction of GNRs with junction states and sizeable large band gaps.
This is demonstrated for an exemplary system in Fig.~\ref{fig_edge_state}.

In conclusion, we predicted the electronic structure, including band gap character and size, and the $\mathbb{Z}_2$ invariant for all geometrically possible cove-edged zigzag graphene nanoribbons, depending on their characteristic structural variables, the width $N$, the distance between coves $a$, and the cove offset $b$. 
The topological properties are impacted by the termination of the ribbons, reflected by the positions of the inversion centers \textit{\textbf{S}} and \textit{\textbf{L}} in the unit cell.
Equations (\ref{eq_rules_Z2_S}) and (\ref{eq_rules_Z2_L}) provide the topological properties, whereas Eq.~(\ref{eq_band_gap}) gives an approximation of the band gap.
We further give rules for constructing GNR junctions with topological edge states.
We thus demonstrate the direct relation between structural and electronic topology in these systems, and we provide simple rules for the design of cove-edged nanoribbons of rich electronic variety, including metals and semiconductors: the latter with a large variation of band gap and with or without topological edge states.
We are confident that similar rules are applicable to other GNR types.

All calculated data are available at the zenodo repository\cite{zenodo}.


\begin{acknowledgments}
This project has been funded by the Deutsche Forschungsgemeinschaft within the Priority Program PP 2244 ``2DMP'' and CRC 1415.
The authors would like to thank Dr. Ji Ma, Dr. Yubin Fu, and Professor Dr. Xinliang Feng for introducing us to the intriguing field of cove-edged zigzag graphene nanoribbons. 
Dr. Thomas Brumme is thanked for helpful discussions.

F. M. Arnold and T-J. Liu contributed equally to this work.
\end{acknowledgments}


%

\clearpage
\onecolumngrid
\flushleft
\appendix*
\section{Supplemental Material}
\renewcommand{\thetable}{S\arabic{table}}
\renewcommand{\thefigure}{S\arabic{figure}}
\renewcommand{\thesection}{S\arabic{section}}
\setcounter{figure}{0} 

\begin{figure}[ht!]
    \centering
    \includegraphics[width=0.9\textwidth]{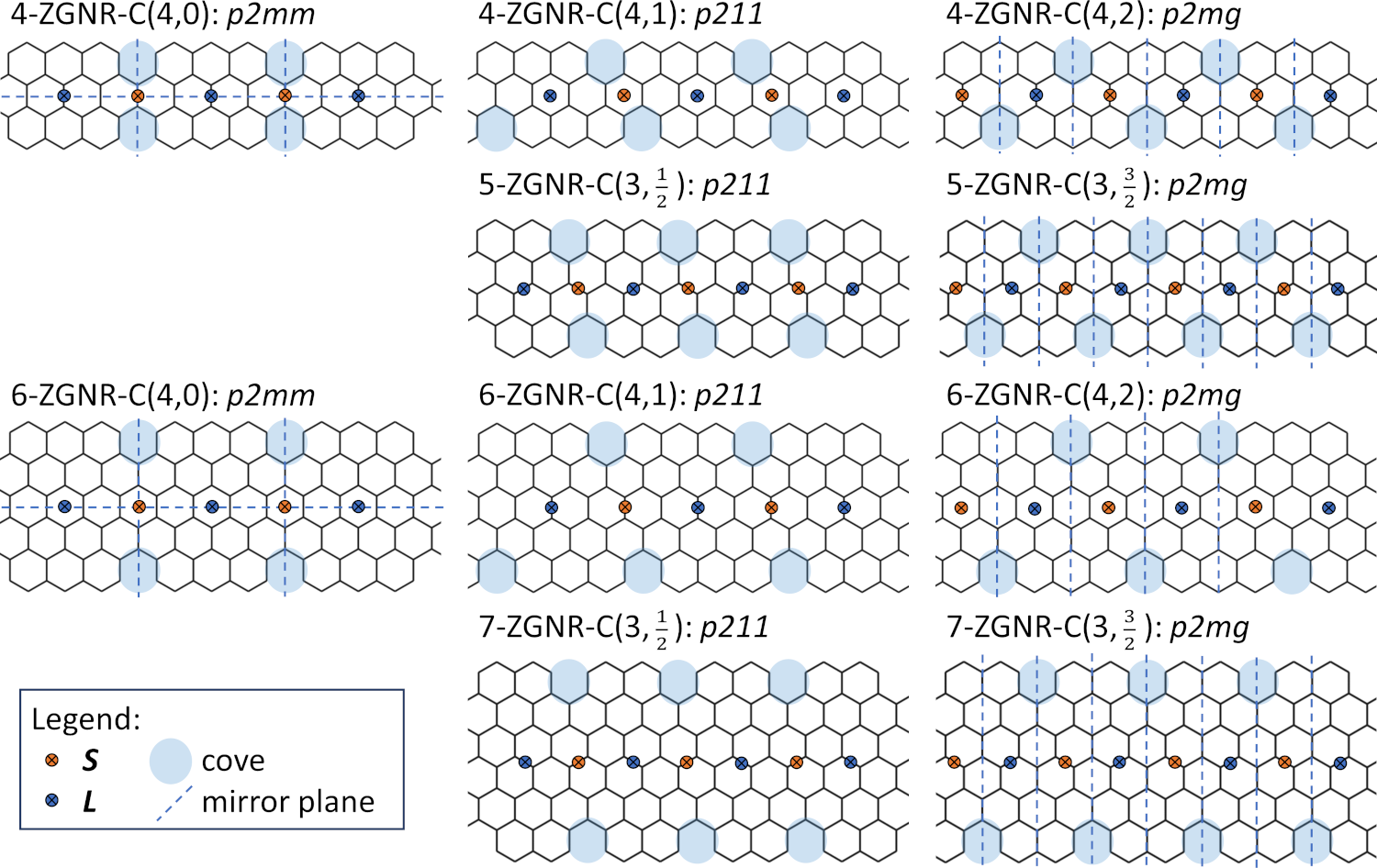}
    \caption{Representative ZGNR-C structures with widths $N=4\dots 7$. The respective label, the Frieze group and the relevant symmetry elements are indicated for each structure.}
    \label{fig_SI_structures}
\end{figure}

\begin{figure}[ht!]
    \centering
    \includegraphics[width=0.6\textwidth]{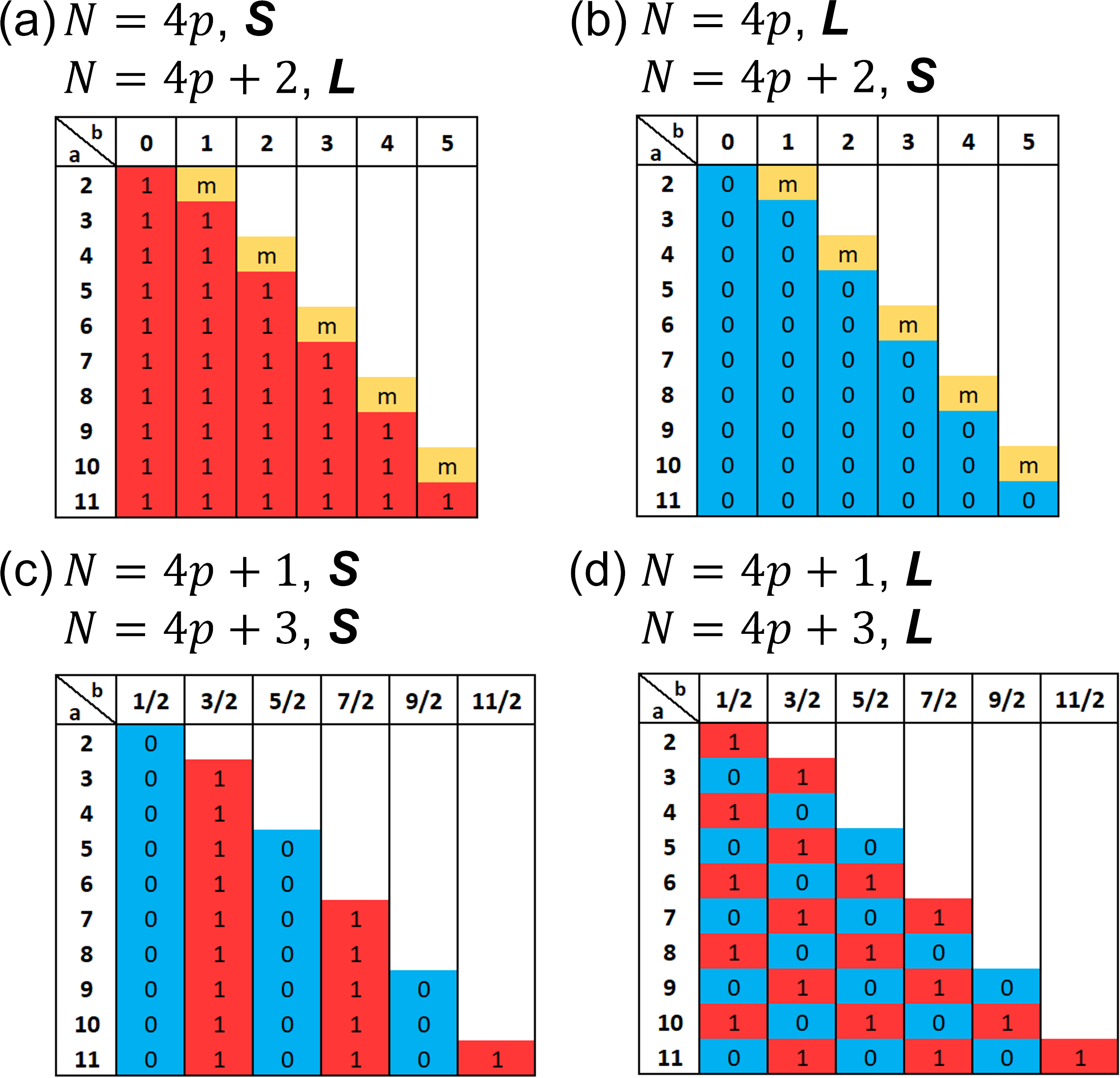}
    \caption{$\mathbb{Z}_2$ invariant for ZGNR-C: (a) $N=4p$ (with integer $p = 1,2,\ldots$) with \textit{\textbf{S}} or $N=4p+2$ with \textit{\textbf{L}} at the boundary, (b) $N=4p$ with \textit{\textbf{L}} or $N=4p+2$ with \textit{\textbf{S}} at the boundary, (c) $N=4p+1$ and $N=4p+3$ with \textit{\textbf{S}} and (d) $N=4p+1$ and $N=4p+3$ with \textit{\textbf{L}} at the boundary. Structure parameters $a$ and $b$ are varied in rows and columns, respectively, with $a\leq11$. Topological insulators ($\mathbb{Z}_2=1$) are marked in red, trivial semiconductors ($\mathbb{Z}_2=0$) in blue, metallic ribbons are indicated as "m" (yellow). }
    \label{fig_SI_Z2_rules}
\end{figure}

\begin{figure}[ht!]
   \centering
    \includegraphics[width=0.9\textwidth]{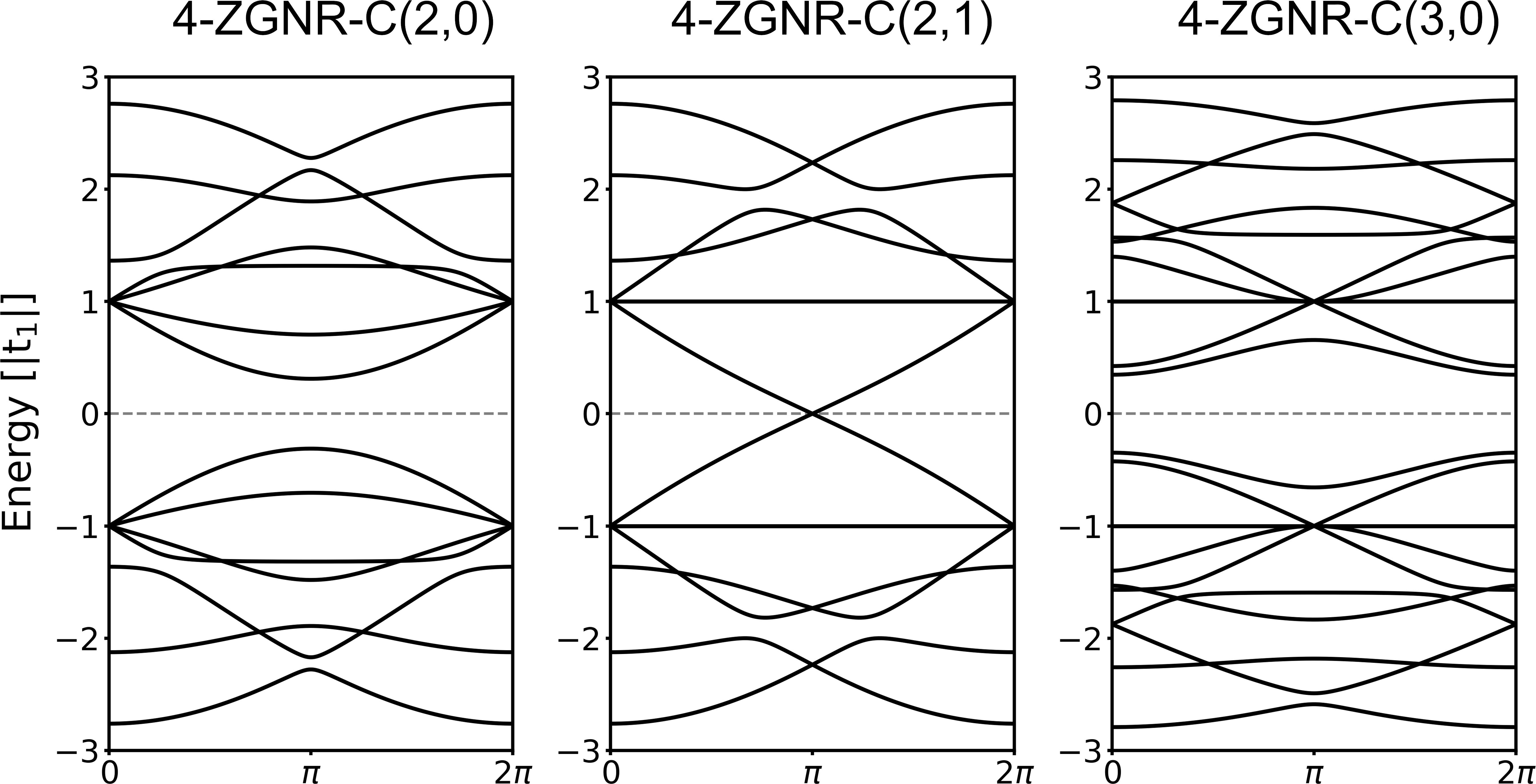}
    \caption{Electronic band structures for 4-ZGNR-C(2,0), 4-ZGNR-C(2,1), and 4-ZGNR-C(3,0) calculated by TB.}
    \label{fig_SI_band}
\end{figure}

\begin{figure}[ht!]
   \centering
    \includegraphics[width=\textwidth]{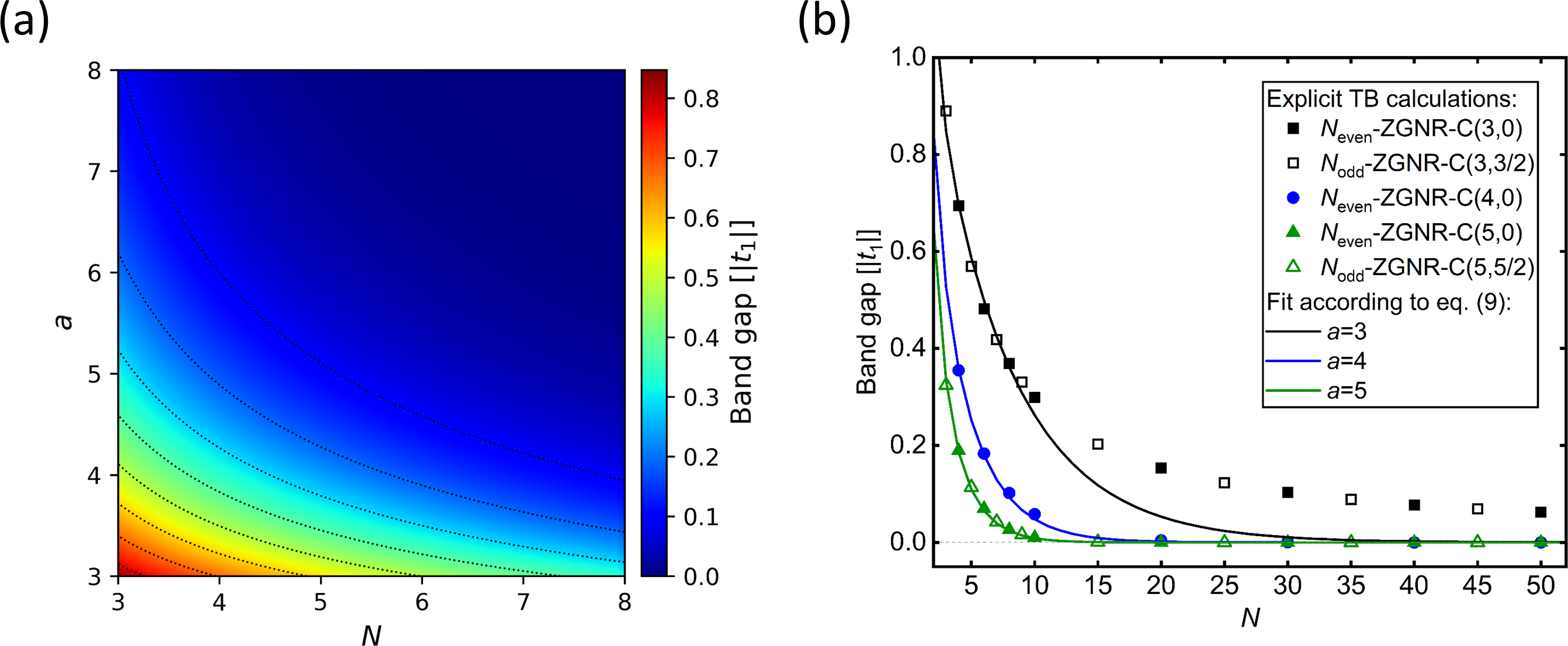}
    \caption{(a) Estimated $\Delta_\mathrm{max}$ as function of $N$ and $a$ according to Eq.~(\ref{eq_Delta_max}). Dashed lines are isolines of $\Delta_\mathrm{max}$ with 0.1 $|t_1|$ difference between each other. (b) Electronic band gap for large $N$, demonstrated using exemplary structures and curves obtained for the chosen $a$ using Eq.~(\ref{eq_Delta_max}).}
    \label{fig_SI_gap}
\end{figure}

\begin{figure}[ht!]
    \centering
    \includegraphics[width=0.6\textwidth]{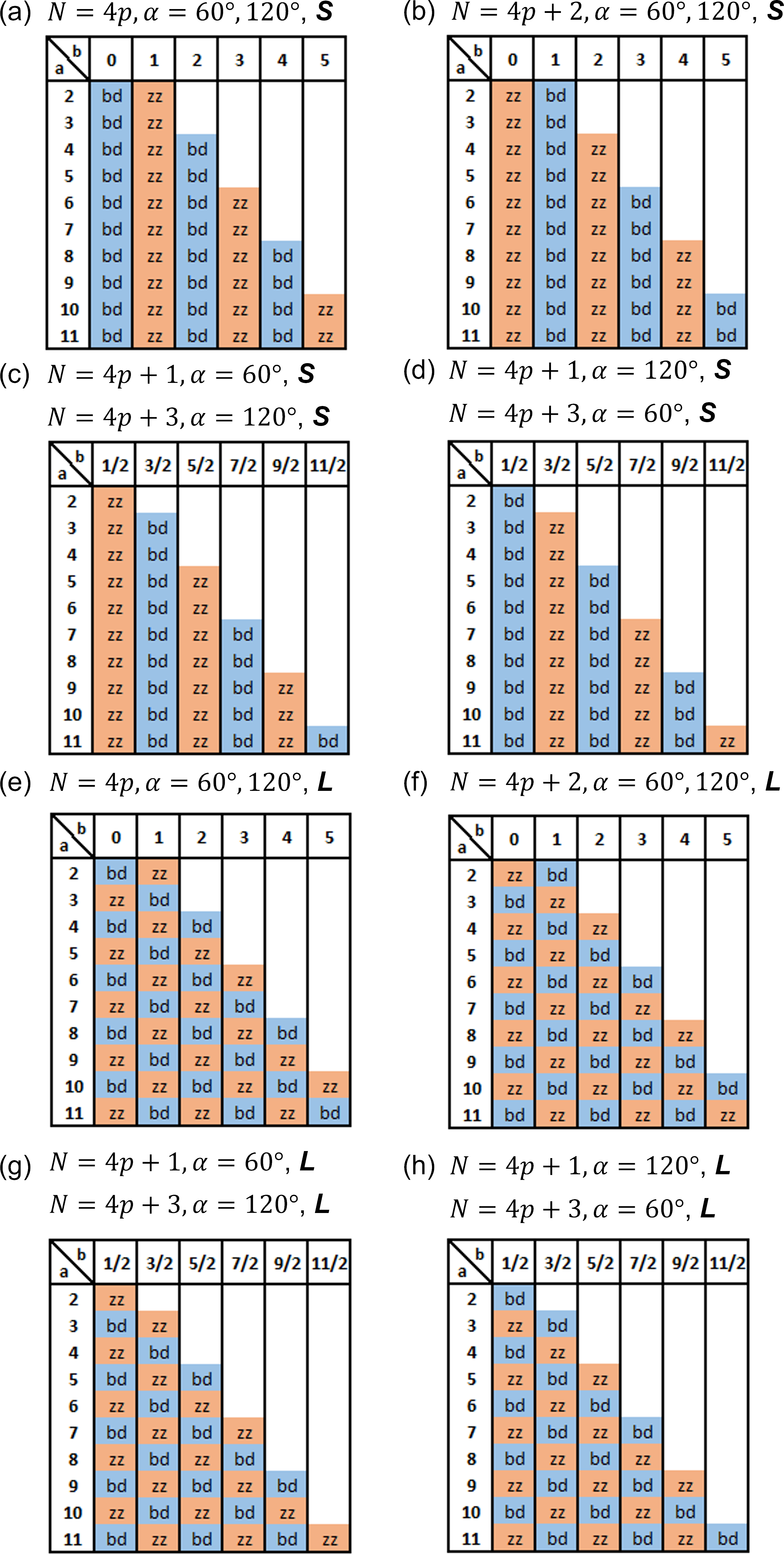}
    \caption{Termination types for ZGNR-C: (a-d) Unit cells with \textit{\textbf{S}} at the boundary with (a) $N=4p$ and (b) $N=4p+2$ for both $\alpha=60^\circ$ and $\alpha=120^\circ$, (c) $N=4p+1, \alpha=60^\circ$ or $N=4p+3, \alpha=120^\circ$ and (d) $N=4p+1, \alpha=120^\circ$ or $N=4p+3, \alpha=60^\circ$. (e-h) Unit cells with \textit{\textbf{L}} at the boundary with (e) $N=4p$ and (f) $N=4p+2$ for both $\alpha=60^\circ$ and $\alpha=120^\circ$, (g) $N=4p+1, \alpha=60^\circ$ or $N=4p+3, \alpha=120^\circ$ and (h) $N=4p+1, \alpha=120^\circ$ or $N=4p+3, \alpha=60^\circ$. Zigzag terminations are indicated by labels "zz" in orange, and bearded by "bd" in blue.}
    \label{fig_SI_termination_rules}
\end{figure}

\begin{sidewaysfigure}[ht!]
   \centering
    \includegraphics[width=\textwidth]{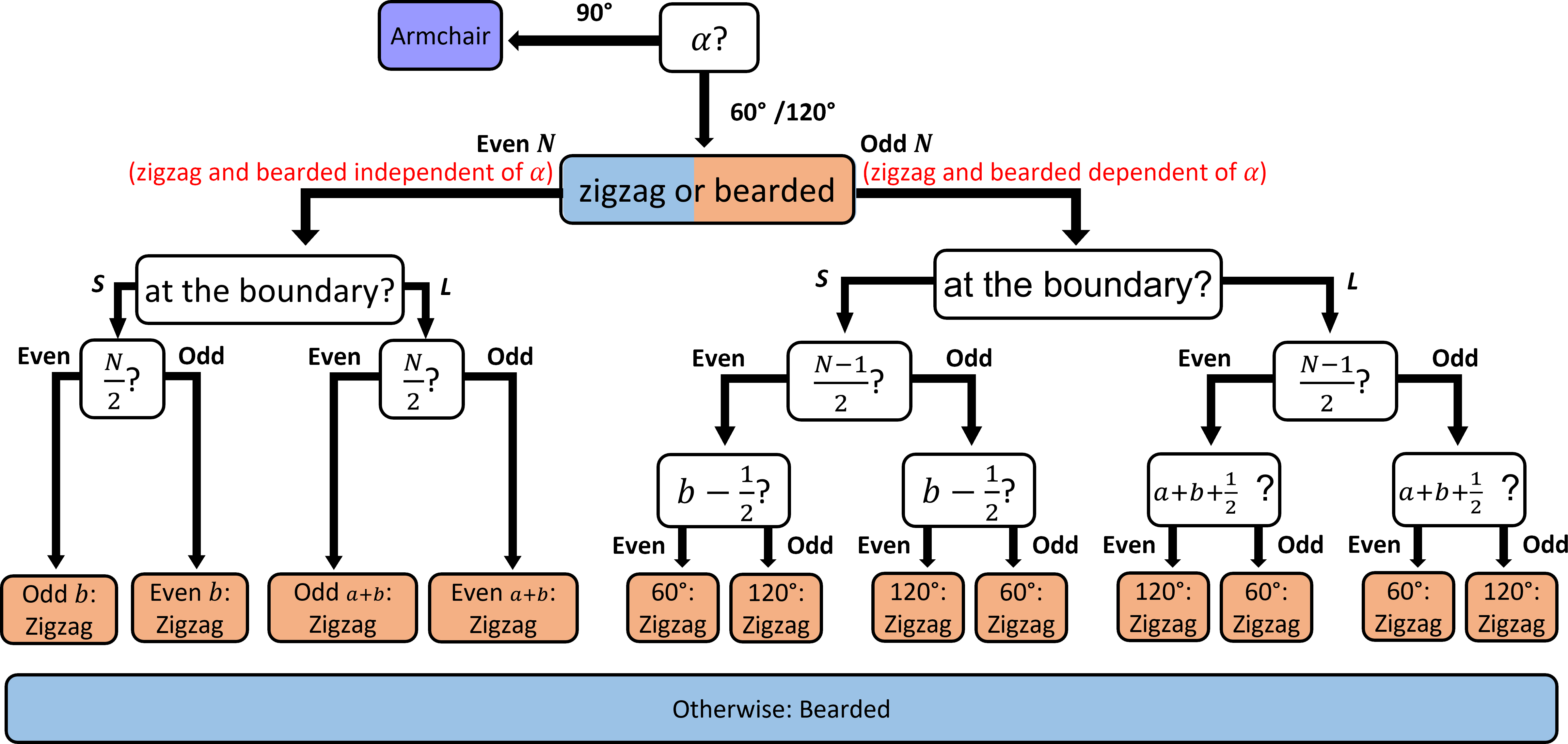}
    \caption{Scheme for realized termination types in ZGNR-C.}
    \label{fig_SI_termination_scheme_full}
\end{sidewaysfigure}

This supplemental material contains: 
\begin{itemize}
    \item Fig.~\ref{fig_SI_structures}, showing ZGNR-C with widths $N=4\dots 7$ as examples for the nomenclature of cove-edged ZGNR. For each structure the symmetry elements and the Frieze group are included.
    \item Section~\ref{SI_sec_method}, detail information on tight-binding calculations.
    \item Fig.~\ref{fig_SI_Z2_rules}, giving tables of $\mathbb{Z}_2$ for an extended set of ZGNR-C, similar to Fig.~\ref{fig_Z2_rules}(a-d).
    \item Fig.~\ref{fig_SI_band}, showing a set of exemplary band structures with a direct band gap at $k=\pi$ if $a$ is even (4-ZGNR-C(2,0)), at $k=0$ if $a$ is odd (4-ZGNR-C(3,0)), or metallic if $b=\frac{a}{2}$.
    \item Fig.~\ref{fig_SI_gap}, visualizing the fit results given in Eq.~(\ref{eq_Delta_max}) and demonstrating the behavior of the band gap for large $N$, testing the validity of the fit for $\Delta_\mathrm{max}$.
    \item Fig.~\ref{fig_SI_termination_rules}, giving tables of realized termination types for an extended set of ZGNR-C with both \textit{\textbf{S}} and \textit{\textbf{L}} as reference points at the boundary, similar to Fig.~\ref{fig_termination_rules}.
    \item Fig.~\ref{fig_SI_termination_scheme_full}, containing a scheme for the realized termination types of ZGNR-C.
\end{itemize}

\clearpage

\smallskip 

\section{Tight-Binding (TB) Calculation Details}
\label{SI_sec_method}
Throughout this paper, the pythTB package, which can be downloaded from \\http://www.physics.rutgers.edu/pythtb/, is used. 
In this package, $\verb|pythtb.tb_model|$ class is the main tight-binding model class.
Dimensionality, orbitals, lattice vectors, on-site energies and hopping parameters can be set up with this class.
Here, only one $\pi$-electron per carbon atom is considered with 1\textsuperscript{st}-neighbor interaction $t_1=-1$.
To satisfy the symmetry requirement, real-space origin is placed at the center of the cell, which is also an inversion center for ZGNR-C.
In this class, band structure can be calculated with given path and k-grid.

$\verb|pythtb.wf_array|$ is a class designed for calculating topological properties.
With $\verb|solve_on_grid|$ function solving the TB model at each grid point within the BZ, and $\verb|berry_phase|$ function doing the numerical integration for all occupied bands, Zak phase is obtained for 1D system.

\newpage

\begin{thebibliography}{31}%
\makeatletter
\providecommand \@ifxundefined [1]{%
 \@ifx{#1\undefined}
}%
\providecommand \@ifnum [1]{%
 \ifnum #1\expandafter \@firstoftwo
 \else \expandafter \@secondoftwo
 \fi
}%
\providecommand \@ifx [1]{%
 \ifx #1\expandafter \@firstoftwo
 \else \expandafter \@secondoftwo
 \fi
}%
\providecommand \natexlab [1]{#1}%
\providecommand \enquote  [1]{``#1''}%
\providecommand \bibnamefont  [1]{#1}%
\providecommand \bibfnamefont [1]{#1}%
\providecommand \citenamefont [1]{#1}%
\providecommand \href@noop [0]{\@secondoftwo}%
\providecommand \href [0]{\begingroup \@sanitize@url \@href}%
\providecommand \@href[1]{\@@startlink{#1}\@@href}%
\providecommand \@@href[1]{\endgroup#1\@@endlink}%
\providecommand \@sanitize@url [0]{\catcode `\\12\catcode `\$12\catcode
  `\&12\catcode `\#12\catcode `\^12\catcode `\_12\catcode `\%12\relax}%
\providecommand \@@startlink[1]{}%
\providecommand \@@endlink[0]{}%
\providecommand \url  [0]{\begingroup\@sanitize@url \@url }%
\providecommand \@url [1]{\endgroup\@href {#1}{\urlprefix }}%
\providecommand \urlprefix  [0]{URL }%
\providecommand \Eprint [0]{\href }%
\providecommand \doibase [0]{http://dx.doi.org/}%
\providecommand \selectlanguage [0]{\@gobble}%
\providecommand \bibinfo  [0]{\@secondoftwo}%
\providecommand \bibfield  [0]{\@secondoftwo}%
\providecommand \translation [1]{[#1]}%
\providecommand \BibitemOpen [0]{}%
\providecommand \bibitemStop [0]{}%
\providecommand \bibitemNoStop [0]{.\EOS\space}%
\providecommand \EOS [0]{\spacefactor3000\relax}%
\providecommand \BibitemShut  [1]{\csname bibitem#1\endcsname}%
\let\auto@bib@innerbib\@empty
\bibitem [{\citenamefont {Ruffieux}\ \emph {et~al.}(2016)\citenamefont
  {Ruffieux}, \citenamefont {Wang}, \citenamefont {Yang}, \citenamefont
  {S{\'a}nchez-S{\'a}nchez}, \citenamefont {Liu}, \citenamefont {Dienel},
  \citenamefont {Talirz}, \citenamefont {Shinde}, \citenamefont {Pignedoli},
  \citenamefont {Passerone} \emph {et~al.}}]{ruffieux2016surface}%
  \BibitemOpen
  \bibfield  {author} {\bibinfo {author} {\bibfnamefont {P.}~\bibnamefont
  {Ruffieux}}, \bibinfo {author} {\bibfnamefont {S.}~\bibnamefont {Wang}},
  \bibinfo {author} {\bibfnamefont {B.}~\bibnamefont {Yang}}, \bibinfo {author}
  {\bibfnamefont {C.}~\bibnamefont {S{\'a}nchez-S{\'a}nchez}}, \bibinfo
  {author} {\bibfnamefont {J.}~\bibnamefont {Liu}}, \bibinfo {author}
  {\bibfnamefont {T.}~\bibnamefont {Dienel}}, \bibinfo {author} {\bibfnamefont
  {L.}~\bibnamefont {Talirz}}, \bibinfo {author} {\bibfnamefont
  {P.}~\bibnamefont {Shinde}}, \bibinfo {author} {\bibfnamefont {C.~A.}\
  \bibnamefont {Pignedoli}}, \bibinfo {author} {\bibfnamefont {D.}~\bibnamefont
  {Passerone}},  \emph {et~al.},\ }\href {\doibase 10.1038/nature17151}
  {\bibfield  {journal} {\bibinfo  {journal} {Nature}\ }\textbf {\bibinfo
  {volume} {531}},\ \bibinfo {pages} {489} (\bibinfo {year}
  {2016})}\BibitemShut {NoStop}%
\bibitem [{\citenamefont {Houtsma}\ \emph {et~al.}(2021)\citenamefont
  {Houtsma}, \citenamefont {de~la Rie},\ and\ \citenamefont
  {St{\"o}hr}}]{houtsma2021atomically}%
  \BibitemOpen
  \bibfield  {author} {\bibinfo {author} {\bibfnamefont {R.~K.}\ \bibnamefont
  {Houtsma}}, \bibinfo {author} {\bibfnamefont {J.}~\bibnamefont {de~la Rie}},
  \ and\ \bibinfo {author} {\bibfnamefont {M.}~\bibnamefont {St{\"o}hr}},\
  }\href {\doibase 10.1039/D0CS01541E} {\bibfield  {journal} {\bibinfo
  {journal} {Chem. Soc. Rev.}\ }\textbf {\bibinfo {volume} {50}},\ \bibinfo
  {pages} {6541} (\bibinfo {year} {2021})}\BibitemShut {NoStop}%
\bibitem [{\citenamefont {Yoon}\ and\ \citenamefont
  {Dong}(2020)}]{yoon2020liquid}%
  \BibitemOpen
  \bibfield  {author} {\bibinfo {author} {\bibfnamefont {K.-Y.}\ \bibnamefont
  {Yoon}}\ and\ \bibinfo {author} {\bibfnamefont {G.}~\bibnamefont {Dong}},\
  }\href {\doibase 10.1039/C9QM00519F} {\bibfield  {journal} {\bibinfo
  {journal} {Mater. Chem. Front.}\ }\textbf {\bibinfo {volume} {4}},\ \bibinfo
  {pages} {29} (\bibinfo {year} {2020})}\BibitemShut {NoStop}%
\bibitem [{\citenamefont {Narita}\ \emph {et~al.}(2019)\citenamefont {Narita},
  \citenamefont {Chen}, \citenamefont {Chen},\ and\ \citenamefont
  {M{\"u}llen}}]{narita2019solution}%
  \BibitemOpen
  \bibfield  {author} {\bibinfo {author} {\bibfnamefont {A.}~\bibnamefont
  {Narita}}, \bibinfo {author} {\bibfnamefont {Z.}~\bibnamefont {Chen}},
  \bibinfo {author} {\bibfnamefont {Q.}~\bibnamefont {Chen}}, \ and\ \bibinfo
  {author} {\bibfnamefont {K.}~\bibnamefont {M{\"u}llen}},\ }\href@noop {}
  {\bibfield  {journal} {\bibinfo  {journal} {Chem. Sci.}\ }\textbf {\bibinfo
  {volume} {10}},\ \bibinfo {pages} {964} (\bibinfo {year} {2019})}\BibitemShut
  {NoStop}%
\bibitem [{\citenamefont {Gr{\"o}ning}\ \emph {et~al.}(2018)\citenamefont
  {Gr{\"o}ning}, \citenamefont {Wang}, \citenamefont {Yao}, \citenamefont
  {Pignedoli}, \citenamefont {Borin~Barin}, \citenamefont {Daniels},
  \citenamefont {Cupo}, \citenamefont {Meunier}, \citenamefont {Feng},
  \citenamefont {Narita} \emph {et~al.}}]{groning2018engineering}%
  \BibitemOpen
  \bibfield  {author} {\bibinfo {author} {\bibfnamefont {O.}~\bibnamefont
  {Gr{\"o}ning}}, \bibinfo {author} {\bibfnamefont {S.}~\bibnamefont {Wang}},
  \bibinfo {author} {\bibfnamefont {X.}~\bibnamefont {Yao}}, \bibinfo {author}
  {\bibfnamefont {C.~A.}\ \bibnamefont {Pignedoli}}, \bibinfo {author}
  {\bibfnamefont {G.}~\bibnamefont {Borin~Barin}}, \bibinfo {author}
  {\bibfnamefont {C.}~\bibnamefont {Daniels}}, \bibinfo {author} {\bibfnamefont
  {A.}~\bibnamefont {Cupo}}, \bibinfo {author} {\bibfnamefont {V.}~\bibnamefont
  {Meunier}}, \bibinfo {author} {\bibfnamefont {X.}~\bibnamefont {Feng}},
  \bibinfo {author} {\bibfnamefont {A.}~\bibnamefont {Narita}},  \emph
  {et~al.},\ }\href {\doibase 10.1038/s41586-018-0375-9} {\bibfield  {journal}
  {\bibinfo  {journal} {Nature}\ }\textbf {\bibinfo {volume} {560}},\ \bibinfo
  {pages} {209} (\bibinfo {year} {2018})}\BibitemShut {NoStop}%
\bibitem [{\citenamefont {Rizzo}\ \emph {et~al.}(2018)\citenamefont {Rizzo},
  \citenamefont {Veber}, \citenamefont {Cao}, \citenamefont {Bronner},
  \citenamefont {Chen}, \citenamefont {Zhao}, \citenamefont {Rodriguez},
  \citenamefont {Louie}, \citenamefont {Crommie},\ and\ \citenamefont
  {Fischer}}]{rizzo2018topological}%
  \BibitemOpen
  \bibfield  {author} {\bibinfo {author} {\bibfnamefont {D.~J.}\ \bibnamefont
  {Rizzo}}, \bibinfo {author} {\bibfnamefont {G.}~\bibnamefont {Veber}},
  \bibinfo {author} {\bibfnamefont {T.}~\bibnamefont {Cao}}, \bibinfo {author}
  {\bibfnamefont {C.}~\bibnamefont {Bronner}}, \bibinfo {author} {\bibfnamefont
  {T.}~\bibnamefont {Chen}}, \bibinfo {author} {\bibfnamefont {F.}~\bibnamefont
  {Zhao}}, \bibinfo {author} {\bibfnamefont {H.}~\bibnamefont {Rodriguez}},
  \bibinfo {author} {\bibfnamefont {S.~G.}\ \bibnamefont {Louie}}, \bibinfo
  {author} {\bibfnamefont {M.~F.}\ \bibnamefont {Crommie}}, \ and\ \bibinfo
  {author} {\bibfnamefont {F.~R.}\ \bibnamefont {Fischer}},\ }\href {\doibase
  10.1038/s41586-018-0376-8} {\bibfield  {journal} {\bibinfo  {journal}
  {Nature}\ }\textbf {\bibinfo {volume} {560}},\ \bibinfo {pages} {204}
  (\bibinfo {year} {2018})}\BibitemShut {NoStop}%
\bibitem [{\citenamefont {Wang}\ \emph {et~al.}(2021)\citenamefont {Wang},
  \citenamefont {Ma}, \citenamefont {Zheng}, \citenamefont {Osella},
  \citenamefont {Arisnabarreta}, \citenamefont {Droste}, \citenamefont {Serra},
  \citenamefont {Ivasenko}, \citenamefont {Lucotti}, \citenamefont {Beljonne}
  \emph {et~al.}}]{wang2021coves}%
  \BibitemOpen
  \bibfield  {author} {\bibinfo {author} {\bibfnamefont {X.}~\bibnamefont
  {Wang}}, \bibinfo {author} {\bibfnamefont {J.}~\bibnamefont {Ma}}, \bibinfo
  {author} {\bibfnamefont {W.}~\bibnamefont {Zheng}}, \bibinfo {author}
  {\bibfnamefont {S.}~\bibnamefont {Osella}}, \bibinfo {author} {\bibfnamefont
  {N.}~\bibnamefont {Arisnabarreta}}, \bibinfo {author} {\bibfnamefont
  {J.}~\bibnamefont {Droste}}, \bibinfo {author} {\bibfnamefont
  {G.}~\bibnamefont {Serra}}, \bibinfo {author} {\bibfnamefont
  {O.}~\bibnamefont {Ivasenko}}, \bibinfo {author} {\bibfnamefont
  {A.}~\bibnamefont {Lucotti}}, \bibinfo {author} {\bibfnamefont
  {D.}~\bibnamefont {Beljonne}},  \emph {et~al.},\ }\href@noop {} {\bibfield
  {journal} {\bibinfo  {journal} {J. Am. Chem. Soc.}\ }\textbf {\bibinfo
  {volume} {144}},\ \bibinfo {pages} {228} (\bibinfo {year}
  {2021})}\BibitemShut {NoStop}%
\bibitem [{\citenamefont {Shinde}\ \emph {et~al.}(2021)\citenamefont {Shinde},
  \citenamefont {Liu}, \citenamefont {Dienel}, \citenamefont {Gr{\"o}ning},
  \citenamefont {Dumslaff}, \citenamefont {M{\"u}hlinghaus}, \citenamefont
  {Narita}, \citenamefont {M{\"u}llen}, \citenamefont {Pignedoli},
  \citenamefont {Fasel} \emph {et~al.}}]{shinde2021graphene}%
  \BibitemOpen
  \bibfield  {author} {\bibinfo {author} {\bibfnamefont {P.~P.}\ \bibnamefont
  {Shinde}}, \bibinfo {author} {\bibfnamefont {J.}~\bibnamefont {Liu}},
  \bibinfo {author} {\bibfnamefont {T.}~\bibnamefont {Dienel}}, \bibinfo
  {author} {\bibfnamefont {O.}~\bibnamefont {Gr{\"o}ning}}, \bibinfo {author}
  {\bibfnamefont {T.}~\bibnamefont {Dumslaff}}, \bibinfo {author}
  {\bibfnamefont {M.}~\bibnamefont {M{\"u}hlinghaus}}, \bibinfo {author}
  {\bibfnamefont {A.}~\bibnamefont {Narita}}, \bibinfo {author} {\bibfnamefont
  {K.}~\bibnamefont {M{\"u}llen}}, \bibinfo {author} {\bibfnamefont {C.~A.}\
  \bibnamefont {Pignedoli}}, \bibinfo {author} {\bibfnamefont {R.}~\bibnamefont
  {Fasel}},  \emph {et~al.},\ }\href {\doibase 10.1016/j.carbon.2020.12.069}
  {\bibfield  {journal} {\bibinfo  {journal} {Carbon}\ }\textbf {\bibinfo
  {volume} {175}},\ \bibinfo {pages} {50} (\bibinfo {year} {2021})}\BibitemShut
  {NoStop}%
\bibitem [{\citenamefont {Yao}\ \emph {et~al.}(2021)\citenamefont {Yao},
  \citenamefont {Zheng}, \citenamefont {Osella}, \citenamefont {Qiu},
  \citenamefont {Fu}, \citenamefont {Schollmeyer}, \citenamefont {M\"uller},
  \citenamefont {Beljonne}, \citenamefont {Bonn}, \citenamefont {Wang} \emph
  {et~al.}}]{yao2021synthesis}%
  \BibitemOpen
  \bibfield  {author} {\bibinfo {author} {\bibfnamefont {X.}~\bibnamefont
  {Yao}}, \bibinfo {author} {\bibfnamefont {W.}~\bibnamefont {Zheng}}, \bibinfo
  {author} {\bibfnamefont {S.}~\bibnamefont {Osella}}, \bibinfo {author}
  {\bibfnamefont {Z.}~\bibnamefont {Qiu}}, \bibinfo {author} {\bibfnamefont
  {S.}~\bibnamefont {Fu}}, \bibinfo {author} {\bibfnamefont {D.}~\bibnamefont
  {Schollmeyer}}, \bibinfo {author} {\bibfnamefont {B.}~\bibnamefont
  {M\"uller}}, \bibinfo {author} {\bibfnamefont {D.}~\bibnamefont {Beljonne}},
  \bibinfo {author} {\bibfnamefont {M.}~\bibnamefont {Bonn}}, \bibinfo {author}
  {\bibfnamefont {H.~I.}\ \bibnamefont {Wang}},  \emph {et~al.},\ }\href
  {\doibase 10.1021/jacs.1c01882} {\bibfield  {journal} {\bibinfo  {journal}
  {J. Am. Chem. Soc.}\ }\textbf {\bibinfo {volume} {143}},\ \bibinfo {pages}
  {5654} (\bibinfo {year} {2021})}\BibitemShut {NoStop}%
\bibitem [{\citenamefont {Li}\ \emph {et~al.}(2021)\citenamefont {Li},
  \citenamefont {Sanz}, \citenamefont {Merino-D{\'\i}ez}, \citenamefont
  {Vilas-Varela}, \citenamefont {Garcia-Lekue}, \citenamefont {Corso},
  \citenamefont {de~Oteyza}, \citenamefont {Frederiksen}, \citenamefont
  {Pe{\~n}a},\ and\ \citenamefont {Pascual}}]{li2021topological}%
  \BibitemOpen
  \bibfield  {author} {\bibinfo {author} {\bibfnamefont {J.}~\bibnamefont
  {Li}}, \bibinfo {author} {\bibfnamefont {S.}~\bibnamefont {Sanz}}, \bibinfo
  {author} {\bibfnamefont {N.}~\bibnamefont {Merino-D{\'\i}ez}}, \bibinfo
  {author} {\bibfnamefont {M.}~\bibnamefont {Vilas-Varela}}, \bibinfo {author}
  {\bibfnamefont {A.}~\bibnamefont {Garcia-Lekue}}, \bibinfo {author}
  {\bibfnamefont {M.}~\bibnamefont {Corso}}, \bibinfo {author} {\bibfnamefont
  {D.~G.}\ \bibnamefont {de~Oteyza}}, \bibinfo {author} {\bibfnamefont
  {T.}~\bibnamefont {Frederiksen}}, \bibinfo {author} {\bibfnamefont
  {D.}~\bibnamefont {Pe{\~n}a}}, \ and\ \bibinfo {author} {\bibfnamefont
  {J.~I.}\ \bibnamefont {Pascual}},\ }\href {\doibase
  10.1038/s41467-021-25688-z} {\bibfield  {journal} {\bibinfo  {journal} {Nat.
  Commun.}\ }\textbf {\bibinfo {volume} {12}},\ \bibinfo {pages} {1} (\bibinfo
  {year} {2021})}\BibitemShut {NoStop}%
\bibitem [{\citenamefont {Hu}\ \emph {et~al.}(2018)\citenamefont {Hu},
  \citenamefont {Xie}, \citenamefont {De~Corato}, \citenamefont {Ruini},
  \citenamefont {Zhao}, \citenamefont {Meggendorfer}, \citenamefont
  {Straas{\o}}, \citenamefont {Rondin}, \citenamefont {Simon}, \citenamefont
  {Li} \emph {et~al.}}]{hu2018bandgap}%
  \BibitemOpen
  \bibfield  {author} {\bibinfo {author} {\bibfnamefont {Y.}~\bibnamefont
  {Hu}}, \bibinfo {author} {\bibfnamefont {P.}~\bibnamefont {Xie}}, \bibinfo
  {author} {\bibfnamefont {M.}~\bibnamefont {De~Corato}}, \bibinfo {author}
  {\bibfnamefont {A.}~\bibnamefont {Ruini}}, \bibinfo {author} {\bibfnamefont
  {S.}~\bibnamefont {Zhao}}, \bibinfo {author} {\bibfnamefont {F.}~\bibnamefont
  {Meggendorfer}}, \bibinfo {author} {\bibfnamefont {L.~A.}\ \bibnamefont
  {Straas{\o}}}, \bibinfo {author} {\bibfnamefont {L.}~\bibnamefont {Rondin}},
  \bibinfo {author} {\bibfnamefont {P.}~\bibnamefont {Simon}}, \bibinfo
  {author} {\bibfnamefont {J.}~\bibnamefont {Li}},  \emph {et~al.},\ }\href
  {\doibase 10.1021/jacs.8b02209} {\bibfield  {journal} {\bibinfo  {journal}
  {J. Am. Chem. Soc.}\ }\textbf {\bibinfo {volume} {140}},\ \bibinfo {pages}
  {7803} (\bibinfo {year} {2018})}\BibitemShut {NoStop}%
\bibitem [{\citenamefont {Cai}\ \emph {et~al.}(2014)\citenamefont {Cai},
  \citenamefont {Pignedoli}, \citenamefont {Talirz}, \citenamefont {Ruffieux},
  \citenamefont {S{\"o}de}, \citenamefont {Liang}, \citenamefont {Meunier},
  \citenamefont {Berger}, \citenamefont {Li}, \citenamefont {Feng} \emph
  {et~al.}}]{cai2014graphene}%
  \BibitemOpen
  \bibfield  {author} {\bibinfo {author} {\bibfnamefont {J.}~\bibnamefont
  {Cai}}, \bibinfo {author} {\bibfnamefont {C.~A.}\ \bibnamefont {Pignedoli}},
  \bibinfo {author} {\bibfnamefont {L.}~\bibnamefont {Talirz}}, \bibinfo
  {author} {\bibfnamefont {P.}~\bibnamefont {Ruffieux}}, \bibinfo {author}
  {\bibfnamefont {H.}~\bibnamefont {S{\"o}de}}, \bibinfo {author}
  {\bibfnamefont {L.}~\bibnamefont {Liang}}, \bibinfo {author} {\bibfnamefont
  {V.}~\bibnamefont {Meunier}}, \bibinfo {author} {\bibfnamefont
  {R.}~\bibnamefont {Berger}}, \bibinfo {author} {\bibfnamefont
  {R.}~\bibnamefont {Li}}, \bibinfo {author} {\bibfnamefont {X.}~\bibnamefont
  {Feng}},  \emph {et~al.},\ }\href {\doibase 10.1038/nnano.2014.184}
  {\bibfield  {journal} {\bibinfo  {journal} {Nat. Nanotechnol.}\ }\textbf
  {\bibinfo {volume} {9}},\ \bibinfo {pages} {896} (\bibinfo {year}
  {2014})}\BibitemShut {NoStop}%
\bibitem [{\citenamefont {Liu}\ \emph {et~al.}(2015)\citenamefont {Liu},
  \citenamefont {Li}, \citenamefont {Tan}, \citenamefont {Giannakopoulos},
  \citenamefont {Sanchez-Sanchez}, \citenamefont {Beljonne}, \citenamefont
  {Ruffieux}, \citenamefont {Fasel}, \citenamefont {Feng},\ and\ \citenamefont
  {M\"ullen}}]{liu2015toward}%
  \BibitemOpen
  \bibfield  {author} {\bibinfo {author} {\bibfnamefont {J.}~\bibnamefont
  {Liu}}, \bibinfo {author} {\bibfnamefont {B.-W.}\ \bibnamefont {Li}},
  \bibinfo {author} {\bibfnamefont {Y.-Z.}\ \bibnamefont {Tan}}, \bibinfo
  {author} {\bibfnamefont {A.}~\bibnamefont {Giannakopoulos}}, \bibinfo
  {author} {\bibfnamefont {C.}~\bibnamefont {Sanchez-Sanchez}}, \bibinfo
  {author} {\bibfnamefont {D.}~\bibnamefont {Beljonne}}, \bibinfo {author}
  {\bibfnamefont {P.}~\bibnamefont {Ruffieux}}, \bibinfo {author}
  {\bibfnamefont {R.}~\bibnamefont {Fasel}}, \bibinfo {author} {\bibfnamefont
  {X.}~\bibnamefont {Feng}}, \ and\ \bibinfo {author} {\bibfnamefont
  {K.}~\bibnamefont {M\"ullen}},\ }\href {\doibase 10.1021/jacs.5b03017}
  {\bibfield  {journal} {\bibinfo  {journal} {J. Am. Chem. Soc.}\ }\textbf
  {\bibinfo {volume} {137}},\ \bibinfo {pages} {6097} (\bibinfo {year}
  {2015})}\BibitemShut {NoStop}%
\bibitem [{\citenamefont {Lee}\ \emph {et~al.}(2018)\citenamefont {Lee},
  \citenamefont {Zhao}, \citenamefont {Cao}, \citenamefont {Ihm},\ and\
  \citenamefont {Louie}}]{lee2018coves}%
  \BibitemOpen
  \bibfield  {author} {\bibinfo {author} {\bibfnamefont {Y.-L.}\ \bibnamefont
  {Lee}}, \bibinfo {author} {\bibfnamefont {F.}~\bibnamefont {Zhao}}, \bibinfo
  {author} {\bibfnamefont {T.}~\bibnamefont {Cao}}, \bibinfo {author}
  {\bibfnamefont {J.}~\bibnamefont {Ihm}}, \ and\ \bibinfo {author}
  {\bibfnamefont {S.~G.}\ \bibnamefont {Louie}},\ }\href@noop {} {\bibfield
  {journal} {\bibinfo  {journal} {Nano Lett.}\ }\textbf {\bibinfo {volume}
  {18}},\ \bibinfo {pages} {7247} (\bibinfo {year} {2018})}\BibitemShut
  {NoStop}%
\bibitem [{\citenamefont {Liu}\ \emph {et~al.}(2020)\citenamefont {Liu},
  \citenamefont {Li}, \citenamefont {Zhang}, \citenamefont {Cao}, \citenamefont
  {Zhu},\ and\ \citenamefont {Shi}}]{liu2020interface}%
  \BibitemOpen
  \bibfield  {author} {\bibinfo {author} {\bibfnamefont {D.-X.}\ \bibnamefont
  {Liu}}, \bibinfo {author} {\bibfnamefont {X.-F.}\ \bibnamefont {Li}},
  \bibinfo {author} {\bibfnamefont {X.-H.}\ \bibnamefont {Zhang}}, \bibinfo
  {author} {\bibfnamefont {X.}~\bibnamefont {Cao}}, \bibinfo {author}
  {\bibfnamefont {X.-J.}\ \bibnamefont {Zhu}}, \ and\ \bibinfo {author}
  {\bibfnamefont {D.}~\bibnamefont {Shi}},\ }\href {\doibase
  10.1021/acs.jpcc.0c04152} {\bibfield  {journal} {\bibinfo  {journal} {J.
  Phys. Chem. C}\ }\textbf {\bibinfo {volume} {124}},\ \bibinfo {pages} {15448}
  (\bibinfo {year} {2020})}\BibitemShut {NoStop}%
\bibitem [{\citenamefont {Hamada}\ \emph {et~al.}(1992)\citenamefont {Hamada},
  \citenamefont {Sawada},\ and\ \citenamefont {Oshiyama}}]{hamada1992new}%
  \BibitemOpen
  \bibfield  {author} {\bibinfo {author} {\bibfnamefont {N.}~\bibnamefont
  {Hamada}}, \bibinfo {author} {\bibfnamefont {S.-i.}\ \bibnamefont {Sawada}},
  \ and\ \bibinfo {author} {\bibfnamefont {A.}~\bibnamefont {Oshiyama}},\
  }\href@noop {} {\bibfield  {journal} {\bibinfo  {journal} {Phys. Rev. Lett.}\
  }\textbf {\bibinfo {volume} {68}},\ \bibinfo {pages} {1579} (\bibinfo {year}
  {1992})}\BibitemShut {NoStop}%
\bibitem [{\citenamefont {Saito}\ \emph
  {et~al.}(1992{\natexlab{a}})\citenamefont {Saito}, \citenamefont {Fujita},
  \citenamefont {Dresselhaus},\ and\ \citenamefont
  {Dresselhaus}}]{saito1992electronic_C60}%
  \BibitemOpen
  \bibfield  {author} {\bibinfo {author} {\bibfnamefont {R.}~\bibnamefont
  {Saito}}, \bibinfo {author} {\bibfnamefont {M.}~\bibnamefont {Fujita}},
  \bibinfo {author} {\bibfnamefont {G.}~\bibnamefont {Dresselhaus}}, \ and\
  \bibinfo {author} {\bibfnamefont {M.~S.}\ \bibnamefont {Dresselhaus}},\
  }\href@noop {} {\bibfield  {journal} {\bibinfo  {journal} {Phys. Rev. B}\
  }\textbf {\bibinfo {volume} {46}},\ \bibinfo {pages} {1804} (\bibinfo {year}
  {1992}{\natexlab{a}})}\BibitemShut {NoStop}%
\bibitem [{\citenamefont {Saito}\ \emph
  {et~al.}(1992{\natexlab{b}})\citenamefont {Saito}, \citenamefont {Fujita},
  \citenamefont {Dresselhaus},\ and\ \citenamefont
  {Dresselhaus}}]{saito1992electronic}%
  \BibitemOpen
  \bibfield  {author} {\bibinfo {author} {\bibfnamefont {R.}~\bibnamefont
  {Saito}}, \bibinfo {author} {\bibfnamefont {M.}~\bibnamefont {Fujita}},
  \bibinfo {author} {\bibfnamefont {G.}~\bibnamefont {Dresselhaus}}, \ and\
  \bibinfo {author} {\bibfnamefont {u.~M.}\ \bibnamefont {Dresselhaus}},\
  }\href@noop {} {\bibfield  {journal} {\bibinfo  {journal} {Appl. Phys.
  Lett.}\ }\textbf {\bibinfo {volume} {60}},\ \bibinfo {pages} {2204} (\bibinfo
  {year} {1992}{\natexlab{b}})}\BibitemShut {NoStop}%
\bibitem [{\citenamefont {Cao}\ \emph {et~al.}(2017)\citenamefont {Cao},
  \citenamefont {Zhao},\ and\ \citenamefont {Louie}}]{cao2017topological}%
  \BibitemOpen
  \bibfield  {author} {\bibinfo {author} {\bibfnamefont {T.}~\bibnamefont
  {Cao}}, \bibinfo {author} {\bibfnamefont {F.}~\bibnamefont {Zhao}}, \ and\
  \bibinfo {author} {\bibfnamefont {S.~G.}\ \bibnamefont {Louie}},\ }\href@noop
  {} {\bibfield  {journal} {\bibinfo  {journal} {Phys. Rev. Lett.}\ }\textbf
  {\bibinfo {volume} {119}},\ \bibinfo {pages} {076401} (\bibinfo {year}
  {2017})}\BibitemShut {NoStop}%
\bibitem [{\citenamefont {Fuess}\ \emph {et~al.}(2010)\citenamefont {Fuess},
  \citenamefont {Hahn}, \citenamefont {Wondratschek}, \citenamefont {Müller},
  \citenamefont {Shmueli}, \citenamefont {Prince}, \citenamefont {Authier},
  \citenamefont {Kopský}, \citenamefont {Litvin}, \citenamefont {Rossmann},
  \citenamefont {Arnold}, \citenamefont {Hall}, \citenamefont {McMahon},
  \citenamefont {Kopský},\ and\ \citenamefont {Litvin}}]{IUC_Frieze}%
  \BibitemOpen
  \bibinfo {editor} {\bibfnamefont {H.}~\bibnamefont {Fuess}}, \bibinfo
  {editor} {\bibfnamefont {T.}~\bibnamefont {Hahn}}, \bibinfo {editor}
  {\bibfnamefont {H.}~\bibnamefont {Wondratschek}}, \bibinfo {editor}
  {\bibfnamefont {U.}~\bibnamefont {Müller}}, \bibinfo {editor} {\bibfnamefont
  {U.}~\bibnamefont {Shmueli}}, \bibinfo {editor} {\bibfnamefont
  {E.}~\bibnamefont {Prince}}, \bibinfo {editor} {\bibfnamefont
  {A.}~\bibnamefont {Authier}}, \bibinfo {editor} {\bibfnamefont
  {V.}~\bibnamefont {Kopský}}, \bibinfo {editor} {\bibfnamefont
  {D.}~\bibnamefont {Litvin}}, \bibinfo {editor} {\bibfnamefont
  {M.}~\bibnamefont {Rossmann}}, \bibinfo {editor} {\bibfnamefont
  {E.}~\bibnamefont {Arnold}}, \bibinfo {editor} {\bibfnamefont
  {S.}~\bibnamefont {Hall}}, \bibinfo {editor} {\bibfnamefont {B.}~\bibnamefont
  {McMahon}}, \bibinfo {editor} {\bibfnamefont {V.}~\bibnamefont {Kopský}}, \
  and\ \bibinfo {editor} {\bibfnamefont {D.}~\bibnamefont {Litvin}},\ eds.,\
  \enquote {\bibinfo {title} {The 7 frieze groups},}\ in\ \href {\doibase
  https://doi.org/10.1107/97809553602060000784} {\emph {\bibinfo {booktitle}
  {International Tables for Crystallography}}}\ (\bibinfo  {publisher} {John
  Wiley \& Sons, Ltd},\ \bibinfo {year} {2010})\ Chap.\ \bibinfo {chapter}
  {2.1}, pp.\ \bibinfo {pages} {31--38}\BibitemShut {NoStop}%
\bibitem [{SM()}]{SM}%
  \BibitemOpen
  \href@noop {} {}\bibinfo {note} {See Supplemental Material at [URL will be
  inserted by publisher] for details on examples of ZGNR-Cs nomenclature, band
  structure, estimated band gaps, as well as extended data sets for
  $\mathbb{Z}_2$ and realized termination of ZGNR-Cs (a scheme is included for
  realized termination of ZGNR-Cs).}\BibitemShut {Stop}%
\bibitem [{\citenamefont {Coh}\ and\ \citenamefont
  {Vanderbilt}(2016)}]{PythTB}%
  \BibitemOpen
  \bibfield  {author} {\bibinfo {author} {\bibfnamefont {S.}~\bibnamefont
  {Coh}}\ and\ \bibinfo {author} {\bibfnamefont {D.}~\bibnamefont
  {Vanderbilt}},\ }\href@noop {} {\enquote {\bibinfo {title} {Pythtb 1.7.2},}\
  }\bibinfo {howpublished} {http://www.physics.rutgers.edu/pythtb/} (\bibinfo
  {year} {2016}),\ \bibinfo {note} {accessed 29 May 2020}\BibitemShut {NoStop}%
\bibitem [{\citenamefont {Zak}(1989)}]{10.1103/PhysRevLett.62.2747}%
  \BibitemOpen
  \bibfield  {author} {\bibinfo {author} {\bibfnamefont {J.}~\bibnamefont
  {Zak}},\ }\href {\doibase 10.1103/PhysRevLett.62.2747} {\bibfield  {journal}
  {\bibinfo  {journal} {Phys. Rev. Lett.}\ }\textbf {\bibinfo {volume} {62}},\
  \bibinfo {pages} {2747} (\bibinfo {year} {1989})}\BibitemShut {NoStop}%
\bibitem [{\citenamefont {Resta}(2000)}]{resta2000manifestations}%
  \BibitemOpen
  \bibfield  {author} {\bibinfo {author} {\bibfnamefont {R.}~\bibnamefont
  {Resta}},\ }\href {\doibase 10.1088/0953-8984/12/9/201} {\bibfield  {journal}
  {\bibinfo  {journal} {J. Phys. Condens. Matter}\ }\textbf {\bibinfo {volume}
  {12}},\ \bibinfo {pages} {R107} (\bibinfo {year} {2000})}\BibitemShut
  {NoStop}%
\bibitem [{\citenamefont {Rhim}\ \emph {et~al.}(2017)\citenamefont {Rhim},
  \citenamefont {Behrends},\ and\ \citenamefont {Bardarson}}]{rhim2017bulk}%
  \BibitemOpen
  \bibfield  {author} {\bibinfo {author} {\bibfnamefont {J.-W.}\ \bibnamefont
  {Rhim}}, \bibinfo {author} {\bibfnamefont {J.}~\bibnamefont {Behrends}}, \
  and\ \bibinfo {author} {\bibfnamefont {J.~H.}\ \bibnamefont {Bardarson}},\
  }\href {\doibase 10.1103/PhysRevB.95.035421} {\bibfield  {journal} {\bibinfo
  {journal} {Phys. Rev. B}\ }\textbf {\bibinfo {volume} {95}},\ \bibinfo
  {pages} {035421} (\bibinfo {year} {2017})}\BibitemShut {NoStop}%
\bibitem [{\citenamefont {Fu}\ and\ \citenamefont
  {Kane}(2007)}]{10.1103/PhysRevB.76.045302}%
  \BibitemOpen
  \bibfield  {author} {\bibinfo {author} {\bibfnamefont {L.}~\bibnamefont
  {Fu}}\ and\ \bibinfo {author} {\bibfnamefont {C.~L.}\ \bibnamefont {Kane}},\
  }\href {\doibase 10.1103/PhysRevB.76.045302} {\bibfield  {journal} {\bibinfo
  {journal} {Phys. Rev. B}\ }\textbf {\bibinfo {volume} {76}},\ \bibinfo
  {pages} {045302} (\bibinfo {year} {2007})}\BibitemShut {NoStop}%
\bibitem [{\citenamefont {Lin}\ and\ \citenamefont
  {Chou}(2018)}]{lin2018topological}%
  \BibitemOpen
  \bibfield  {author} {\bibinfo {author} {\bibfnamefont {K.-S.}\ \bibnamefont
  {Lin}}\ and\ \bibinfo {author} {\bibfnamefont {M.-Y.}\ \bibnamefont {Chou}},\
  }\href@noop {} {\bibfield  {journal} {\bibinfo  {journal} {Nano Lett.}\
  }\textbf {\bibinfo {volume} {18}},\ \bibinfo {pages} {7254} (\bibinfo {year}
  {2018})}\BibitemShut {NoStop}%
\bibitem [{\citenamefont {Jiang}\ and\ \citenamefont
  {Louie}(2020)}]{jiang2020topology}%
  \BibitemOpen
  \bibfield  {author} {\bibinfo {author} {\bibfnamefont {J.}~\bibnamefont
  {Jiang}}\ and\ \bibinfo {author} {\bibfnamefont {S.~G.}\ \bibnamefont
  {Louie}},\ }\href@noop {} {\bibfield  {journal} {\bibinfo  {journal} {Nano
  Lett.}\ }\textbf {\bibinfo {volume} {21}},\ \bibinfo {pages} {197} (\bibinfo
  {year} {2020})}\BibitemShut {NoStop}%
\bibitem [{\citenamefont {Saroka}\ \emph {et~al.}(2014)\citenamefont {Saroka},
  \citenamefont {Batrakov},\ and\ \citenamefont
  {Chernozatonskii}}]{saroka2014edge}%
  \BibitemOpen
  \bibfield  {author} {\bibinfo {author} {\bibfnamefont {V.}~\bibnamefont
  {Saroka}}, \bibinfo {author} {\bibfnamefont {K.}~\bibnamefont {Batrakov}}, \
  and\ \bibinfo {author} {\bibfnamefont {L.}~\bibnamefont {Chernozatonskii}},\
  }\href@noop {} {\bibfield  {journal} {\bibinfo  {journal} {Solid State
  Phys.}\ }\textbf {\bibinfo {volume} {56}},\ \bibinfo {pages} {2135} (\bibinfo
  {year} {2014})}\BibitemShut {NoStop}%
\bibitem [{\citenamefont {Son}\ \emph {et~al.}(2006)\citenamefont {Son},
  \citenamefont {Cohen},\ and\ \citenamefont {Louie}}]{son2006energy}%
  \BibitemOpen
  \bibfield  {author} {\bibinfo {author} {\bibfnamefont {Y.-W.}\ \bibnamefont
  {Son}}, \bibinfo {author} {\bibfnamefont {M.~L.}\ \bibnamefont {Cohen}}, \
  and\ \bibinfo {author} {\bibfnamefont {S.~G.}\ \bibnamefont {Louie}},\ }\href
  {\doibase 10.1103/PhysRevLett.97.216803} {\bibfield  {journal} {\bibinfo
  {journal} {Phys. Rev. Lett.}\ }\textbf {\bibinfo {volume} {97}},\ \bibinfo
  {pages} {216803} (\bibinfo {year} {2006})}\BibitemShut {NoStop}%
\bibitem [{zenodo()}]{zenodo}%
  \BibitemOpen
  \href@noop {} {}\bibinfo {note} {F. Arnold, T.-J. Liu, A. Kuc, and T. Heine, Structure-imposed electronic topology in cove-edged graphene nanoribbons (2022), DOI:10.5281/zenodo.7254203.}\BibitemShut {Stop}%
\end{thebibliography}
\end{document}